\documentclass[prd,amsfonts,aps,nofootinbib,notitlepage,11pt]{revtex4-1}
\pdfoutput=1
\usepackage{amsmath,amssymb}
\usepackage{graphicx}
\usepackage[utf8x]{inputenc}
\usepackage{cancel}
\usepackage[colorlinks=true]{hyperref}
\usepackage{color}
\usepackage{array,multirow} 
\usepackage{float}
\usepackage{ulem}

\renewcommand{\arraystretch}{1.1}

\newcommand{\mx}{m^2_{\psi}}
\newcommand{\Mi}{M^2_{i}}
\newcommand{\Mj}{M^2_{j}}
\newcommand{\m}{m^2}
\newcommand{\mi}{m^2_{i}}
\newcommand{\mj}{m^2_{j}}
\newcommand{\M}{M^2}

\begin{document}
\title{Singlet fermion dark matter and Dirac neutrinos
from Peccei-Quinn symmetry}
\author{Cristian D. R. Carvajal}
\email{cdavid.ruiz@udea.edu.co}
\email{cristianruiz6246@correo.itm.edu.co}
\affiliation{Instituto de F\'isica, Universidad de Antioquia, Calle 70 No. 52-21, Medell\'in, Colombia. \\
Instituto Tecnol\'ogico Metropolitano, Facultad de Ciencias, Medell\'in, Colombia.} 
\author{Robinson Longas}
\email{robinson.longas@udea.edu.co}
\affiliation{Instituto de F\'isica, Universidad de Antioquia, Calle 70 No. 52-21, Medell\'in, Colombia.}
\author{Oscar Rodríguez}
\email{oscara.rodriguez@udea.edu.co}
\affiliation{Instituto de F\'isica, Universidad de Antioquia, Calle 70 No. 52-21, Medell\'in, Colombia.}
\author{\'Oscar Zapata}
\email{oalberto.zapata@udea.edu.co}
\affiliation{Instituto de F\'isica, Universidad de Antioquia, Calle 70 No. 52-21, Medell\'in, Colombia.}
\date{\today}
          
\begin{abstract}
The Peccei-Quinn (PQ) mechanism not only acts as an explanation for the absence of strong CP violation but also can play a main role in the solution to other open questions in particle physics and cosmology.   
Here we consider a model that identifies the PQ symmetry as a common thread in the solution to the strong CP problem, the generation of radiative Dirac neutrino masses and  the origin of a multicomponent dark sector.  
Specifically,  scotogenic neutrino masses arise at one loop level with the lightest fermionic mediator field  acting as the second dark matter (DM) candidate thanks to the residual $Z_2$ symmetry resulting from the PQ symmetry breaking.    
We perform a phenomenological analysis addressing the constraints coming from the direct searches of DM, neutrino oscillation data and charged lepton flavor violating (LFV) processes.
We find that the model can be partially probed in future facilities searching for WIMPs and axions, and accommodates rates for rare leptonic decays that are within the expected sensitivity of upcoming LFV experiments. 
\end{abstract}

\maketitle

\section{Introduction}
The apparent non-observation of CP violation in the QCD Lagrangian represents one of the most active subjects in high energy physics, both theoretical and experimentally speaking. 
In the theory side, the absence of strong CP violation can be dynamically explained invoking the Peccei-Quinn (PQ) mechanism \cite{Peccei:1977hh}, which considers the spontaneous breaking of an anomalous global $U(1)$ symmetry with the associated pseudo-Nambu-Goldstone boson, the (QCD) axion~\cite{Weinberg:1977ma, Wilczek:1977pj}. 
The axion itself turns to be a promising candidate for making up the dark matter (DM) of the Universe thanks to a variety of production mechanisms \cite{Sikivie:2006ni}, for instance via the vacuum misalignment mechanism~\cite{Preskill:1982cy, Abbott:1982af, Dine:1982ah}. 
Besides this, it is remarkable that the physics behind the PQ mechanism can be also used to address other open questions in particle physics and cosmology such as neutrino masses~\cite{Mohapatra:1982tc, Shafi:1984ek, Langacker:1986rj, Shin:1987xc, He:1988dm, Berezhiani:1989fp, Bertolini:1990vz, Ma:2001ac}, baryon asymmetry~\cite{Servant:2014bla, Ipek:2018lhm, Croon:2019ugf, Co:2019wyp} and inflation~\cite{Linde:1991km, Folkerts:2013tua, Fairbairn:2014zta, Ballesteros:2016euj}.  

The recent analysis~\cite{Carvajal:2018ohk} considering the PQ mechanism as the responsible for the massiveness of neutrinos revealed that it is also possible to consistently accommodate radiative Dirac neutrino masses\footnote{See Refs.~\cite{Chen:2012baa,Dasgupta:2013cwa,Bertolini:2014aia,Gu:2016hxh,Ma:2017zyb,Ma:2017vdv,Suematsu:2017kcu,Suematsu:2017hki,Reig:2018ocz,Reig:2018yfd,Peinado:2019mrn,Baek:2019wdn,delaVega:2020jcp,Dias:2020kbj,Baek:2020ovw} for recent and related works.} with a viable WIMP DM candidate, thus providing a set of multicomponent scotogenic models with Dirac neutrinos. 
Concretely, in these scenarios one-loop Dirac neutrino masses are generated through the $d=5$ effective operator $\bar{L}\tilde{H}N_RS$ \cite{Ma:2016mwh,Yao:2018ekp} once the axionic field $S$ develops a vacuum expectation value, with the contributions arising from the tree-level realizations of such an operator being forbidden thanks to the charge assignment.  
As a further consequence of the PQ symmetry, the residual discrete symmetry that is left over renders stable the lightest particle mediating the neutrino masses, and since such a particle must be electrically neutral, it turns out that the setup also accommodates a second DM species~\cite{Baer:2011hx, Bae:2013hma, Dasgupta:2013cwa, Alves:2016bib, Ma:2017zyb, Chatterjee:2018mac}. 

In this work we perform a phenomenological analysis of the T3-1-A-I model introduced in Ref.~\cite{Carvajal:2018ohk}. 
In order to determine the viable parameter space of the model we take into account the constraints coming from direct detection experiments, lepton flavor violation (LFV) processes, DM relic density and neutrino physics. 
We find that for a wide and typical range of the parameter values, the model easily satisfies these constraints and, additionally, future experiments will be able to test a considerable portion of the parameter space. 

The layout of this paper is organized as follows. The main features of the model are presented in Section \ref{sec:model}.
In Section \ref{DirectDetection} we determine the elastic scattering cross section between the WIMP particle and nucleons, and estimate the expected number of events in current and future direct detection experiments.
Section \ref{sec:phenomenology} is dedicated to a numerical analysis addressing the DM and LFV phenomenology. Finally, we conclude in Section \ref{sec:summary}. 
 
\section{The model}
\label{sec:model}
\begin{table}[t]
  \setlength{\tabcolsep}{5pt} \global\long\def\arraystretch{1.1}
  \centering
  \begin{tabular}{|c|c|c|c|c|c|c|c|c|c|}
    \hline
    & $L_{\beta}$ & $\ell_{R\beta}$ & $N_{R\beta}$ & $S$  &  $\psi_i$  & $\Phi$    & $H_2$       & $D_L$   & $D_R$\\
    \hline
    \hline
    $U(1)_L$  & $1$ & $1$   & $1$   & $0$  &  $1$         & $0$      & $0$         & $0$    & $0$ \\
    \hline
    $U(1)_\textrm{PQ}$  & $2$ & $2$   & $0$   & $2$  &  $3$         & $3$      & $1$         & $1$    & $-1$ \\
    \hline
    $Z_2$     & $+$ & $+$   & $+$   & $+$  & $-$         & $-$      & $-$         & $-$    & $-$ \\
    \hline   
  \end{tabular}
  \caption{Lepton number and PQ charge assignments for the model particles. Here, both the SM Higgs, $H_1$, and the ordinary quarks are even under the discrete symmetry and neutral under the global ones. The transformation properties under the remnant $Z_2$ symmetry are also shown. $\beta$ is a family index ($\beta =1,2,3)$ and $i=1,2$.}
  \label{tab:chargesT31}
\end{table}

As usual in models with massive Dirac neutrinos, this model extends the SM with three singlet Weyl fermions $N_{R\beta}\,(\beta=1,2,3)$ $-$the right-handed partners of the SM neutrinos. 
The one-loop neutrino mass generation additionally demands \cite{Carvajal:2018ohk} the introduction of one $SU(2)_L$ doublet scalar $H_2$, one singlet scalar $\Phi$ and two singlet Dirac fermions $\psi_i\,(i=1,2)$. As the last piece we have a chiral exotic down-type quark $D$ which is added in order to realize the hadronic KSVZ-type axion model~\cite{Kim:1979if,Shifman:1979if}. 
In Table \ref{tab:chargesT31} are displayed the charge assignments under the $U(1)_{L}$ and $U(1)_{PQ}$ global symmetries, as well as under the remnant $Z_2$ symmetry.   
Notice that under this discrete symmetry the mediator fields in the neutrino mass diagram are odd,  which implies that the lightest of them can be considered as a (WIMP-like) DM candidate\footnote{We assume that the lightest particle charged under the PQ symmetry is electrically neutral.}. 

The relevant part of the scalar potential can be expressed as
\begin{align}
   \mathcal{V}&\supset-\mu_1^2|H_1|^2+\lambda_1|H_1|^4+\mu^2_S|S|^2+\lambda_S|S|^4+
   \mu^2_2|H_2|^2+\lambda_2|H_2|^4+\lambda_3 |H_1|^2|H_2|^2+\lambda_4 |H_1^\dagger H_2|^2\nonumber\\
   &\,+\mu^2_\Phi|\Phi|^2+\lambda_\Phi|\Phi|^4+\lambda_{h\Phi} |H_1|^2|\Phi|^2
+ \lambda'\left( \Phi^* S  \tilde{H}_2^\dagger  \tilde{H}_1 +\textrm{h.c.}\right),
 \end{align}
where the coupling constants associated to the quartic interaction terms  $|\Phi|^2|S|^2$, $|H_2|^2|S|^2$, $|H_2|^2|\Phi|^2$ and $|H_1|^2|S|^2$ have been assumed to be small. Since the $(H_1^\dagger H_2)^2+(H_2^\dagger H_1)^2$ term is forbidden, it follows that the neutral component of $H_2$ remains as a complex field and does not get splitted into a CP-even and a CP-odd field. Nevertheless, it does get mixed with $\Phi$ through the term proportional to $\lambda'$ (since both scalar fields do not acquire a nonzero vacuum expectation value).    
We parametrize the scalar fields as
\begin{align}
    S=\frac{1}{\sqrt{2}}\left(\rho + v_S\right)e^{ia/v_S},\hspace{0.5cm}
    H_1=\begin{pmatrix}
    0\\
    \frac{v_{\textrm{SM}}+h}{\sqrt{2}}
    \end{pmatrix},\hspace{0.5cm}
    H_2 =\begin{pmatrix}
    H^+\\
    H^0
\end{pmatrix},
\end{align}
where $\rho$ stands for the radial component of the field $S$ whose mass is set by the scale of the PQ symmetry breaking $v_S$, whereas 
the angular part of $S$ corresponds to the QCD axion $a$, $h$ is the SM Higgs boson and $v_{\textrm{SM}}=246.22$ GeV.

In the basis $(H^0,\Phi)$, the mass matrix for the $Z_2$-odd neutral scalars 
\begin{equation}\label{Meven}
 \mathcal{M}_{0}=
 \left( 
  \begin{array}{cc}
    \mu_2^2+\frac{1}{2}(\lambda_3+\lambda_4)v_{\textrm{SM}}^2 & \frac{1}{2}\lambda' v_Sv_{\textrm{SM}}\\
   \frac{1}{2} \lambda'v_Sv_{\textrm{SM}}& \mu_\Phi^2+\frac{1}{2}\lambda_{h\Phi}v_{\textrm{SM}}^2
  \end{array}
 \right ) , 
\end{equation}
leads to the mass eigenstates $S_{1,2}$ (both with two degrees of freedom since they are complex) via the transformation 
\begin{align}
H^0 &=\cos\varphi\ S_1 + \sin\varphi\ S_2 = \sum_{j=1,2} C_{H j} S_j\equiv C_{H j} S_j ,\label{H02}\\
\Phi &=-\sin\varphi\ S_1 + \cos\varphi\ S_2 = \sum_{j=1,2} C_{\Phi j} S_j \equiv C_{\Phi j} S_j, \label{Phi}
\end{align}
where we have defined $C_{H 1}=\cos\varphi,\, C_{H 2}=\sin\varphi,\, C_{\Phi 1}=-\sin\varphi$ and $C_{\Phi 2}=\cos\varphi$. The mixing angle is defined through the expression\footnote{  
A tiny value for $\lambda'$ is not only necessary to reproduce the observed neutrino phenomenology but also to have the complex scalars $S_{1,2}$ at or below the TeV mass scale. 
This is in consonance with the requirement of demanding a tiny value  for the scalar couplings between the axion field $S$ and the other scalar fields, as happens in most of the axion models.}
\begin{align}
    \sin(2\varphi)=\frac{\lambda' v_{\textrm{SM}} v_S}{m^2_{S_{2}}-m^2_{S_{1}}},
\end{align}
being $m_{S_{1,2}}$ the eigenvalues of $\mathcal{M}_0$. In our analysis we will take $m_{S_1} < m_{S_2}$, which implies that for $\varphi \to 0$ the heavier scalar is mainly singlet, whereas for $\varphi\to \pm\, \pi/2$ the heavier one is mainly doublet. For the charged scalar $H^\pm$ we find that its mass is given by $m^2_{H^\pm}=\mu_2^2+\frac{1}{2}\lambda_3v_{\textrm{SM}}^2$, just as happens in the inert doublet model.

\subsection{Neutrino masses and charged LFV}\label{sec:neutrinos}
\begin{figure}
  \centering
  \includegraphics[scale=0.5]{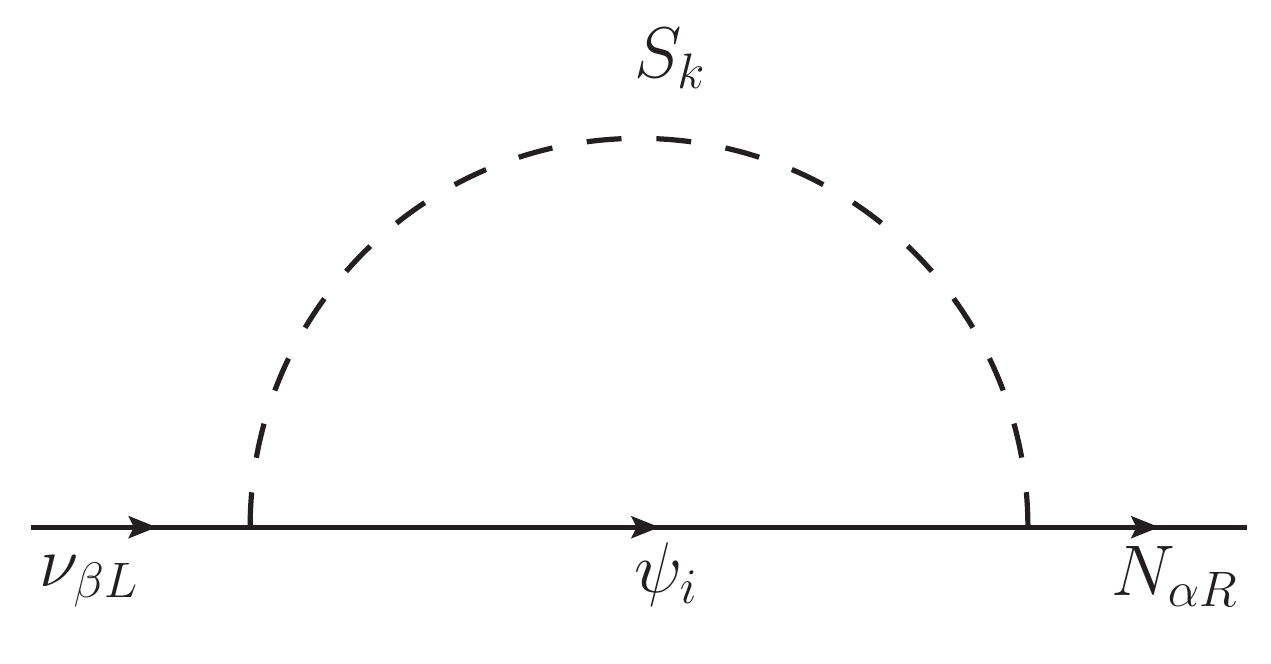}
  \caption{One-loop Feynman diagram leading to Dirac neutrino masses. The loop mediators are the singlet fermions $\psi_{1,2}$ and the neutral scalars $S_{1,2}$.}
  \label{NeutrinoMassDiagram}
\end{figure}

The new Yukawa interactions involving the SM leptons are given by 
\begin{align}\label{Yukawa Lagrangian}
   \mathcal{L}_Y&= y_{i\beta}\overline{\psi_i}\tilde{H}_2^\dagger L_{\beta}+h_{\beta i}\Phi^*\overline{N_{R\beta}} \psi_i+\textrm{h.c.},
 \end{align}
 where $y_{i\beta}$ and $h_{\beta i}$ are $2\times 3$ and $3\times 2$ Yukawa matrices, respectively. 
After the spontaneous symmetry breaking, scotogenic Dirac neutrino masses are generated at one loop \cite{Farzan:2012sa,Ma:2016mwh,Wang:2016lve,Borah:2016zbd,Wang:2017mcy,Yao:2017vtm,Yao:2018ekp,Ma:2017kgb,Wang:2017mcy,Reig:2018mdk, Calle:2018ovc, Calle:2019mxn, Bernal:2021ezl}  as is illustrated in Fig. \ref{NeutrinoMassDiagram}.  The effective mass matrix can be written as 
\begin{equation}\label{neutrinomasses1}
  \big(M_\nu\big)_{\alpha\beta}=\sum_ih_{\alpha i}y_{ i\beta}\kappa_i,
\end{equation}
where 
\begin{equation}\label{neutrinomasses2}
    \kappa_i=\frac{\sin(2\varphi)}{32\pi^2}m_{\psi_i}\left[F\left(\frac{m_{S_{2}}^2}{m_{\psi_i}^2}\right)-F\left(\frac{m_{S_{1}}^2}{m_{\psi_i}^2}\right)\right],
\end{equation}
and $F(x)=x\ln(x)/(x-1)$. 
The Dirac neutrino mass matrix $M_\nu$ can be diagonalized through $M_\nu=VM^{diag}_{\nu}U^\dagger$, where $U$ and $V$ are unitary matrices and $M^{diag}_{\nu}$ is a diagonal mass matrix containing, in general, three mass eigenvalues different from zero. However, due to the flavor structure of $M_{\nu}$ only two neutrinos are massive (det$(M_{\nu})=0$). In the basis where the charged lepton mass matrix is diagonal  the $U$ unitary matrix can be identified with the PMNS matrix~\cite{Zyla:2020zbs}, whereas $V$ can be assumed diagonal without loss of generality. 
This allows us to express the $y$ Yukawa couplings in terms of the $h$ ones. In the case of the normal neutrino mass hierarchy (NH)
\begin{align}\label{eq:yukawasNH}
    y_{1\beta} &=  \frac{ h_{32}m_2 U^*_{\beta 2} -  h_{22}m_3 U^*_{\beta 3}}{(h_{21}h_{32}-h_{31}h_{22})k_1}\,,\notag\\ 
    y_{2\beta} &=  \frac{h_{21} m_3 U^*_{\beta 3} -  h_{31}m_2 U^*_{\beta 2}}{(h_{21}h_{32}-h_{31}h_{22})k_2}\,,
\end{align}
with $h_{11}=h_{12}=0$, whereas in the inverted hierarchy (IH) case  
\begin{align}\label{eq:yukawasIH}
    y_{1\beta} &= \frac{h_{12}m_2 U^*_{\beta 2}-h_{22}m_3 U^*_{\beta 1}}{(h_{12}h_{21}-h_{11}h_{22})k_1}\,,\notag\\
    y_{2\beta} &= \frac{h_{21}m_3 U^*_{\beta 1}-h_{11}m_2 U^*_{\beta 2}}{(h_{12}h_{21}-h_{11}h_{22})k_2}\,,
\end{align}
with $h_{31}=h_{32}=0$. Notice that one of the right-handed neutrinos becomes decoupled because we are considering a scenario with the minimal set of singlet fermions\footnote{In the scenario with three singlet fermions, all the neutrino eigenstates would be massive.}.

\begin{figure}[t!]
  \includegraphics[scale=0.9]{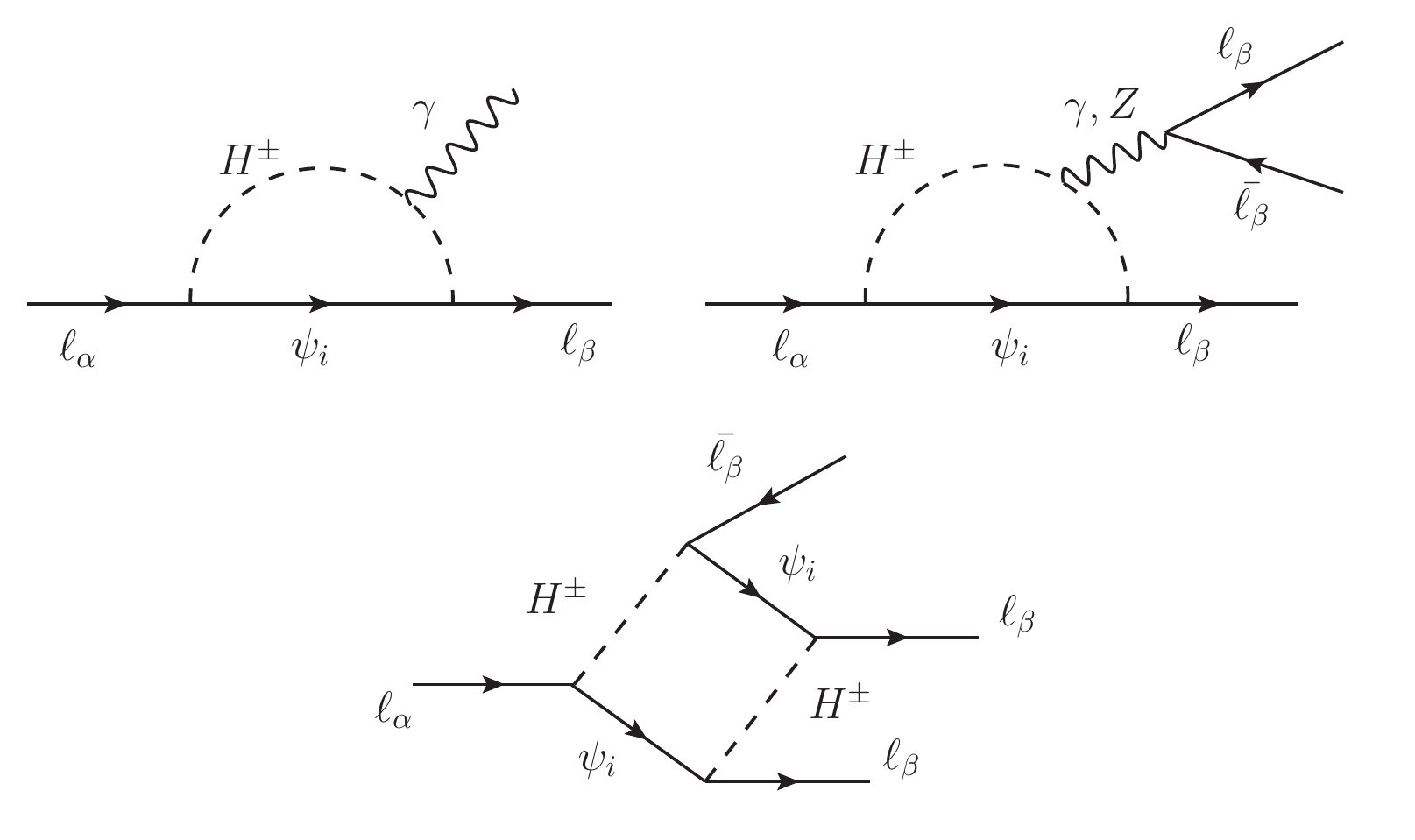}
  \caption{Feynman diagrams for the $\ell_\alpha\rightarrow \ell_{\beta}\gamma$ and $\ell_{\alpha} \rightarrow \ell_{\beta}\bar{\ell}_{\beta}\ell_{\beta}$ charged LFV processes present in the model.}
  \label{fig:LFVdiagrams}
\end{figure}

Although Diracness of neutrino masses is compatible with the conservation of the total lepton number, family lepton number violation is unavoidable due to neutrino oscillations. 
In this model,  LFV processes involving charged leptons such as $\ell_{\alpha} \rightarrow \ell_{\beta}\gamma$, $\ell_{\alpha} \rightarrow 3\ell_{\beta}$ and $\mu-e$ conversion in nuclei are induced at one-loop level, involving only $y_{i\beta}$ Yukawa interactions and mediated by the $H^{\pm}$ charged scalar and the $\psi_{i}$ neutral fermions. 
The decay rate for the $\ell_{\alpha} \rightarrow \ell_{\beta}\gamma$ processes (see the top left panel of Fig. \ref{fig:LFVdiagrams}) neglecting the lepton masses at the final states is given by 
\begin{equation}
 \Gamma\big(\ell_\alpha\rightarrow \ell_{\beta}\gamma\big)=\frac{e^2m_\alpha^5}{16^3\pi^5m^4_{H^\pm}}\sum_i\left|y_{i\alpha}y_{i\beta}^*\right|^2\left[\frac{2t_i^2+5t_i-1}{12(t_i-1)^3}-\frac{t_i^2\log t_i}{2(t_i-1)^4}\right]^2\,,
\end{equation}
where $t_i\equiv m_{\psi_i}^2/m_{H^\pm}^2$. 
Concerning the $\ell_{\alpha} \rightarrow \ell_{\beta}\bar{\ell}_{\beta}\ell_{\beta}$ processes, there are two kinds of diagrams (see Fig.~\ref{fig:LFVdiagrams}): the $\gamma$- and $Z$- penguin diagrams (top-right panel) and the box diagrams (bottom panel). The contribution from the Higgs-penguin diagrams is suppressed for the first two charged leptons generations due to their small Yukawa couplings\footnote{The contribution of those processes involving tau leptons is not negligible, but the corresponding limits are less restrictive.}. It follows that the $\ell_{\alpha} \rightarrow \ell_{\beta}\bar{\ell_{\beta}}\ell_{\beta}$ processes contain four kind of contributions: the photonic monopole, photonic dipole,  $Z$-penguin and box diagrams\footnote{Notice that the Yukawa interactions also lead to neutrino three-body decays $\nu_h\to\nu_l\bar{\nu}_l\nu_l$, with $m_{\nu_h}>m_{\nu_l}$, via a box diagram similar to the bottom panel in figure \ref{fig:LFVdiagrams}, with the charged leptons replaced by neutrinos and the charged scalar $H^\pm$ by the neutral one $H^0$. Since the decay rate for these processes \cite{Arganda:2005ji} is proportional to the ratio $m_{\nu_h}^5/m_{S_1}^4$, the expected lifetime is several orders of magnitude larger than the age of the Universe.} . In contrast, the photonic dipole contribution is the only one present in the $\ell_{\alpha} \rightarrow \ell_{\beta} \gamma$ processes.
Finally, the $\mu-e$ conversion diagrams are obtained when the pair of lepton lines attached to the photon and $Z$ boson in the penguin diagrams (see top panels of Fig.~\ref{fig:LFVdiagrams}) are replaced by a pair of light quark lines\footnote{Higgs-penguin diagrams are again suppressed, in this case by the Yukawa couplings to light quarks.}. For the $\mu-e$ conversion in nuclei there are no box diagrams since the $Z_2$-odd particles do not couple to quarks at tree level. Accordingly, the photonic non-dipole and dipole terms along with the $Z-$penguin one are the only terms that contribute to the $\mu-e$ conversion processes. 
In this work we calculate the rates for $\ell_\alpha \rightarrow \ell_\beta\bar{\ell_\beta}\ell_\beta$ and $\mu-e$ through the chain {\tt SARAH} \cite{Staub:2013tta,Staub:2015kfa}, {\tt SPheno}~\cite{Porod:2003um,Porod:2011nf} and {\tt FlavorKit} \cite{Porod:2014xia}.

\subsection{Two-component dark matter} \label{sec:axionDM}
The natural DM candidate in models featuring a PQ symmetry is the axion itself since the associated energy density decreases as non-relativistic matter does \cite{diCortona:2015ldu,DiLuzio:2020wdo,Arias:2012az,Sikivie:2006ni}.  
The amount of axion relic abundance depends on whether the PQ symmetry is broken before or after inflation. On the  one hand, if PQ symmetry is broken after the inflationary epoch the axion field would be randomly distributed over the whole range $(0,2\pi v_S)$, meaning that the initial misalignment angle $\theta_a$ takes different values in different patches of the Universe resulting in the average $\langle\theta_a\rangle=\pi/\sqrt{3}$. In this case, topological defects such as string axions and domain walls ~\cite{Davis:1985pt, Harari:1987ht, Battye:1994au, Hiramatsu:2012gg} also contribute to the axion relic abundance. 
On the other hand, when PQ symmetry is broken before the inflationary epoch and is not restored during the reheating phase the axion field is uniform over the observable Universe, meaning that the initial misalignment angle takes a single value in the interval $(0,2\pi]$.

For simplicity purposes, we assume that the reheating temperature after inflation is below the PQ symmetry breaking scale, in which case the axion abundance is settled to 
\cite{Abbott:1982af,Bae:2008ue}
\begin{align}
\label{eq:axionDMabunce}
 \Omega_ah^2\approx 0.18\theta_a^2\left(\frac{v_S}{10^{12} \textrm{ GeV}}\right)^{1.19}.
\end{align}
It follows that the axion can be the main DM constituent if $v_S\sim 10^{12}$ GeV for a no fine-tuned $\theta_a$ (that is $\theta_a \approx\mathcal{O}(1)$). Under this premise, the axion window becomes $m_a \sim (1-10)$~$\mu$eV. Nevertheless, the axion can give a subdominant contribution to the relic DM abundance for lower values of $v_S$, thus allowing for a  multicomponent DM scenario.    

In addition to the axion, this model brings along with a second DM candidate since the remnant $Z_2$ symmetry renders stable the lightest particle charged under it.  The case of $S_1$ being that candidate turns to be ruled out since direct detection searches have excluded models where the DM candidate has a direct coupling to the $Z$ gauge boson. Therefore, $\psi_1\equiv\psi$ becomes the second viable DM candidate of the model. 
According to Eq.~(\ref{Yukawa Lagrangian}), $\psi$ only interacts with the SM particles via the Yukawa interactions $y$ and $h$, and since these must be non-negligible in order to explain the neutrino oscillation parameters $\psi$ necessarily reaches thermal equilibrium with the SM plasma. 
The $\psi$ relic abundance is determined by the cross sections of the annihilation and co-annihilation processes $\bar{\psi}\psi\to\bar{\ell}\ell,\bar{\nu}\nu$ and $H^+\psi\to\ell\gamma$, respectively. 
Let us to stress that the $h$ interactions can actually take large values 
because they are not taking part of the LFV processes, which means that $\psi$ may feature a large annihilation cross section. 
If the fermion DM and scalar mediator masses are assumed to be sufficiently non-degenerate, co-annihilations can be neglected, and thus the relic abundance simply depends on the $\psi$-annihilation cross section. 
In our numerical analysis, nonetheless, we use {\tt micrOMEGAs}~\cite{Belanger:2013oya} in order to take into account all the relevant processes that contribute to the setting of the relic abundance of $\psi$. 

\section{Direct detection of fermion dark matter}
\label{DirectDetection}
Being a SM singlet that couples to leptons and $Z_2$-odd scalars, $\psi$ does not have tree-level interactions with the SM quarks. However, the interactions involving $Z_2$-odd particles allow us to  construct effective interactions at one-loop level between singlet fermion DM and quarks.  

\begin{figure}[t]
    \centering
 \includegraphics[scale=0.7]{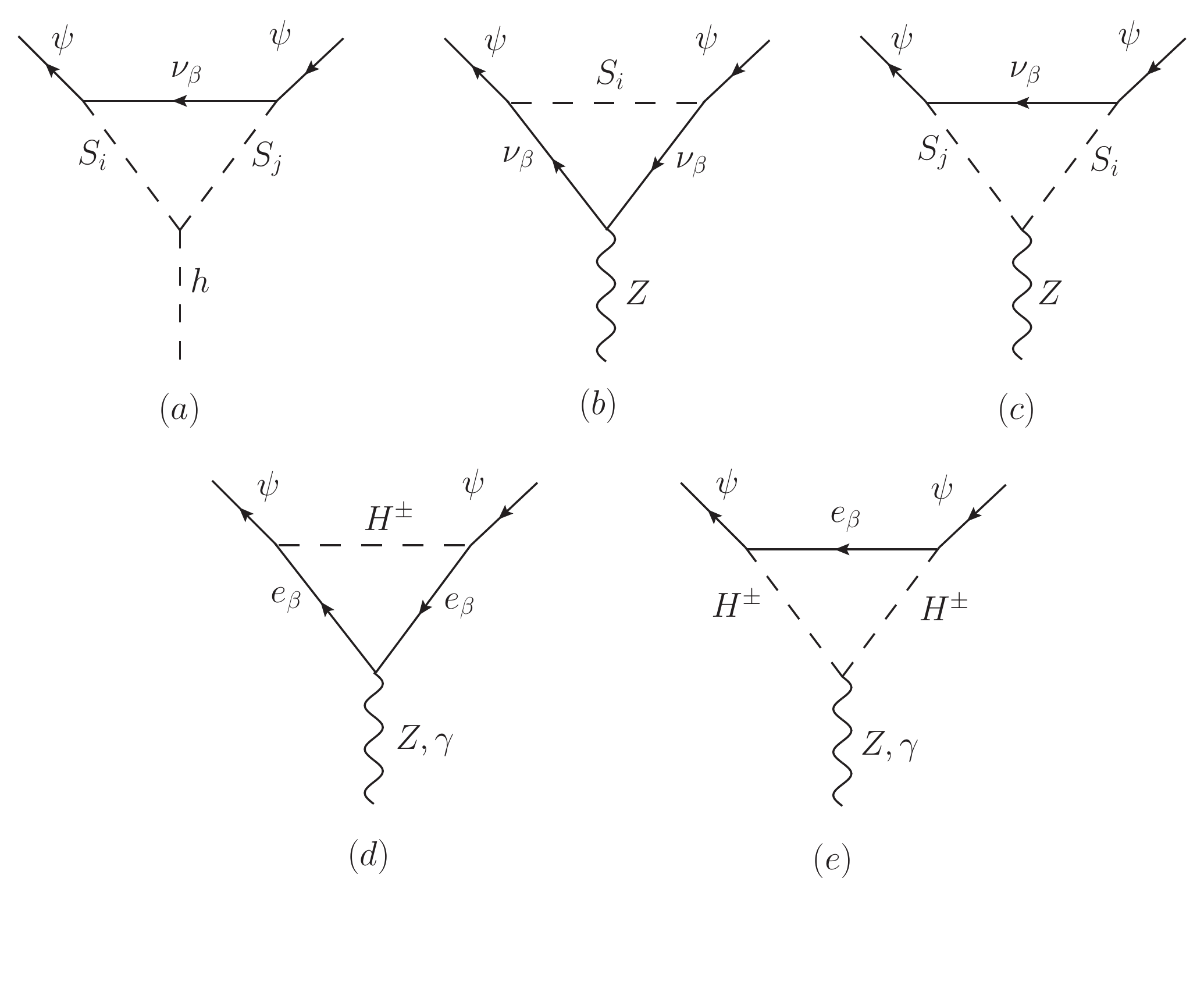}
    \caption{One-loop effective couplings between the fermion DM candidate and the SM neutral bosons.} 
    \label{oneloopdiagrams}
\end{figure}

In the basis of mass eigenstates, the relevant interaction terms involved in the direct detection of $\psi$ are 
\begin{align}
-\mathcal{L} \supset C^{\beta i}_{\psi S\nu} \overline{\psi}_{R} S_i \nu_{\beta L} + C^{\beta}_{\psi H e} \overline{\psi}_{R} H^{\pm} e_{\beta L} + C^{\beta i}_{\psi S N}\overline{\psi}_{L}S_i N_{\beta R} + C^{ij}_{SSh} S^*_i S_j h + \textrm{h.c.},\label{L2}
\end{align}
where a sum over repeated latin and greek indices is implied. The new coefficients in this expression are defined as 
\begin{align}
C^{\beta i}_{\psi S\nu} = y_{1\beta}C_{H i},\ \ C^\beta_{\psi H e} = -y_{1\beta},\ \ C^{\beta i}_{\psi S N} = h_{\beta 1}C_{\Phi i},\ \  C^{ij}_{SSh} = \lambda_{h\Phi} v_H C_{\Phi i}C_{\Phi j} + \frac{\lambda' v_S}{2}C_{\Phi i} C_{H j}.
\end{align}
The interplay of the above interactions with the gauge and scalar interactions lead to the effective one-loop couplings between $\psi$ and the Higgs, photon and $Z$ bosons as shown in Fig. \ref{oneloopdiagrams}. 

The differential spin-independent cross section for the fermion DM particle being scattered by a target nucleus of mass $m_T$ and atomic and mass numbers $A$ and $Z$ can be expressed as \cite{Hisano:2018bpz}
\begin{align}
\frac{d\sigma_{\psi T}}{dE_R}&=F^2_c(E_R)\left\lbrace\frac{Z^2e^2}{4\pi}\left[\frac{1}{E_R}-\frac{1}{E^{max}_R(v^2_{rel})}\right]\left(C^\gamma_M\right)^2
+\frac{Z^2e^2}{4\pi v^2_{rel}}\frac{1}{E_R}\left(C^\gamma_E\right)^2\right.\notag\\
&\left. + \frac{m_T}{2\pi v^2_{rel}}\left|Z\left(f^p_S+f^p_V-eC^\gamma_R-\frac{e}{2m_{\psi}}C^\gamma_M\right)+\left(A-Z\right)\left(f^n_S+f^n_V\right)\right|^2\right\rbrace.\label{DiffCrossSec}
\end{align}
Here $v_{rel}$ is the relative velocity between $\psi$ and the nucleus, and $E_R$ denotes the recoil energy  of the nucleus due to the interaction. The maximum value of $E_R$, $E^{max}_R$, is related with $v_{rel}$ through 
\begin{align}
    E^{max}_R(v^2_{rel})=\frac{2m^2_{\psi}m_Tv^2_{rel}}{(m_{\psi}+m_T)^2}.
\end{align}
The effective couplings $C^\gamma_{M\left(E\right)}$ and $C^\gamma_R$ correspond to the Wilson coefficients describing the interaction with the photon field as a result of the magnetic (electric) dipole moment and the charge radius of the fermion DM. In this model these quantities are given by 
\begin{align}
    C^{\gamma}_{M}&= -\dfrac{e}{2(4\pi)^2m_{\psi}} \vert y_{1\beta}\vert^2\ g_{M1}\left(m_{\psi},m_{H^\pm},m_{e_\beta}\right),\\
    C^{\gamma}_{E}&=  0,\\
    C^{\gamma}_{R}&=-\dfrac{e}{2(4\pi)^2m^2_{\psi}} \vert y_{1\beta}\vert^2\ g_{R1}\left(m_{\psi},m_{H^\pm},m_{e_\beta}\right),
\end{align}
where explicit expressions for the one-loop functions $g_{M1}$ and $g_{R1}$ are reported in Appendix \ref{sec:oneloopfunctions}. The vanishing of $C^\gamma_E$ has its origin in the absence of a coupling between the right-handed electron and $\psi$ as can be seen from Eq.~(\ref{L2}). 
The interaction with nucleons is described by the effective scalar and vector couplings $f^{(N)}_S$ and $f^{(N)}_V$ $\left(N=p,n\right)$, respectively.  The scalar coupling is given in terms of the $\psi$-quark (-gluon) scalar couplings $C^{q(g)}_S$ and the matrix elements $f^{(N)}_{T_q} = \langle N |m_q \overline{q}q|N\rangle/m_N$\footnote{For the numerical analysis shown in Sec. \ref{sec:phenomenology} we considered the values 
\begin{align*}
f^{(p)}_{T_u}&=0.018,\ f^{(p)}_{T_d}=0.027, \ f^{(p)}_{T_s}=0.037,\ f^{(p)}_{T_G} = 1-\sum_{q=u,d,s}f^{(p)}_{T_q}= 0.918,\\
f^{(n)}_{T_u}&=0.013,\ f^{(n)}_{T_d}=0.040, \ f^{(n)}_{T_s}=0.037,\ f^{(n)}_{T_G} = 1-\sum_{q=u,d,s}f^{(n)}_{T_q}= 0.090,
\end{align*}
as suggested by \cite{Ellis:2018dmb}.}.  
From Ref.~\cite{SHIFMAN1978443} it reads 
\begin{align}
f^{(N)}_S &= m_N\left[\sum_{q=u,d,s} C^q_S f^{(N)}_{T_q} -\frac{8}{9}C^g_S f^{(N)}_{T_G}\right],
\end{align}
where $m_N$ is the nucleon mass. On the other hand, the vector coupling can be expressed in terms of the $\psi$-quark vector couplings $C^q_V$ as
\begin{align}
f^{(N)}_V&=\begin{cases}
2C^u_V+C^d_V,\,\,\,\mathrm{for} \ N=p\,,\\ 
C^u_V+2C^d_V,\,\,\,\mathrm{for} \ N=n\,.
\end{cases} 
\end{align}
In this model, the $\psi$ effective interactions with quarks and gluons can be cast as 
\begin{align}
C^q_S &= -4C^g_S=-\frac{1}{\sqrt{2}m^2_h v_{\mathrm{SM}}}C_{\psi h},\\
C^u_V &= -\frac{2}{v^2_{\mathrm{SM}}}\left(\frac{1}{2}-\frac{4}{3}\sin^2\theta_W\right)C_{\psi Z},\\
C^d_V &= -\frac{2}{v^2_{\mathrm{SM}}}\left(-\frac{1}{2}+\frac{2}{3}\sin^2\theta_W\right)C_{\psi Z},
\end{align}
being $C_{\psi h(Z)}$ the effective scalar (vector) coupling between $\psi$ and the Higgs ($Z$) boson. These are given by
\begin{align}
C_{\psi h} & = \frac{\tilde{g}_{h1}(m_{\psi},m_{S_i},m_{S_j},m_{\nu_\beta})}{\sqrt{2}(4\pi)^2 m_{\psi}} C_{\Phi i} \left(\lambda_{h\Phi}v_{\mathrm{SM}}C_{\Phi j}+\frac{\lambda'v_S}{2}C_{H j}\right)\nonumber\\
&\times\left(\vert y_{1\beta}\vert^2C_{Hi}C_{Hj} +\vert h_{\beta 1}\vert^2 C_{\Phi i}C_{\Phi j}\right),\label{Cpsih}\\
C_{\psi Z} &= \frac{\vert y_{1\beta}\vert^2}{2(4\pi)^2}\left\lbrace \left(\sin^2\theta_W-\frac{1}{2}\right)\left[g_{Z1}(m_{\psi},m_{H^\pm},m_{e_\beta},m_{e_\beta})+\tilde{g}_{Z1}(m_{\psi},m_{H^\pm},m_{H^\pm},m_{e_\beta})\right]\right.\notag\\ 
&\left. +\sin^2\theta_W\frac{m^2_{e_\beta}}{m^2_{\psi}} \ g_{Z2}(m_{\psi},m_{H^\pm},m_{e_\beta},m_{e_\beta})+\frac{C^2_{Hi}}{2}\left[g_{Z1}(m_{\psi},m_{S_i},m_{\nu_\beta},m_{\nu_\beta})\right.\right.\notag\\
&\left. +\ C^2_{Hj}\tilde{g}_{Z1}(m_{\psi},m_{S_i},m_{S_j},m_{\nu_\beta})\right]\bigg\} +\frac{\vert h_{\beta 1}\vert^2}{2(4\pi)^2}\frac{C_{Hi}C_{\Phi i}C_{Hj}C_{\Phi j}}{2} \tilde{g}_{Z1}(m_{\psi},m_{S_i},m_{S_j},m_{\nu_\beta}).\label{CpsiZ}
\end{align}
As indicated above, a sum over repeated indices is implied and the definitions of the one-loop functions $\tilde{g}_{h1}$, $g_{Z1}$, $\tilde{g}_{Z1}$ and $g_{Z2}$ are reported in Appendix \ref{sec:oneloopfunctions}. 
By last, the recoil-energy dependent nuclear form factor $F^2_c(E_R)$ in Eq.~(\ref{DiffCrossSec}) reads~\cite{Lewin:1995rx,Helm:1956zz} 
\begin{align}
    F^2_c(E_R)=\left[3\frac{j_1(qR)}{qR}\right]^2e^{-q^2s^2}, 
\end{align}
where $j_1$ is the spherical Bessel function of the first kind, $q=\sqrt{2m_TE_R}$ and $R = \sqrt{c^2+\frac{7}{3}\pi^2a^2-5s^2}$ with $c = (1.23 A^{1/3}-0.6)$ fm, $a=0.52$ fm and $s = 0.9$ fm.

In order to estimate the expected number of $\psi$-nuclei scattering events in a direct detection experiment like XENON1T \cite{Aprile:2017iyp}, we calculate the differential event rate per unit of detector mass through \cite{Hisano:2018bpz}
\begin{align}
    \frac{dR}{dE_R}=\frac{\rho_{DM}}{m_Tm_{\psi}}\int_{v_{min}(E_R)}^\infty d^3\textbf{v}_{rel} v_{rel} f_{\oplus}(\textbf{v}_{rel})\frac{d\sigma_{\psi T}}{dE_R}.
\end{align}
Here $\rho_{DM}\approx 0.3$ GeV/cm$^{3}$ is the local DM density,  $v_{min}(E_R)$ is the minimum speed needed to yield a recoil with energy $E_R$, which can be determined from
\begin{align}
    v_{min}(E_R)=\sqrt{\frac{(m_{\psi}+m_T)^2E_R}{2m^2_{\psi}m_T}},
\end{align}
and $f_{\oplus}(\textbf{v}_{rel})$ stands for the DM velocity distribution measured with respect to the lab frame. With respect to the galactic frame, this distribution is assumed to follow a Maxwell-Boltzmann one, \textit{i.e.}  
\begin{align}
    f(\mathbf{v})=\begin{cases}\frac{1}{N}e^{-|\mathbf{v}|^2/v^2_0},\ \text{for}\ \ |\mathbf{v}|<v_{\mathrm{esc}},\\
    0,\ \ \text{for}\ \ |\mathbf{v}|>v_{\mathrm{esc}}, \end{cases}
\end{align}
where the maximum speed is equal to the galaxy escape velocity, $v_{\mathrm{esc}}$, and 
\begin{align}
    N=\pi^{3/2}v^3_0\left[ \mathrm{erf}\left(\dfrac{v_{\mathrm{esc}}}{v_0}\right)-\dfrac{2v_{\mathrm{esc}}}{\sqrt{\pi}v_0}e^{-\left(\frac{v_{\mathrm{esc}}}{v_0}\right)^2} \right].
\end{align}
In this way, if $\textbf{v}_E$ is the velocity of the Earth with respect to the galactic frame, then\footnote{Given the functional dependence of $\frac{d\sigma_{\psi T}}{dE_R}$ with $v_{rel}$, the integrals  
\begin{align}
    \zeta(E_R)=\int_{v_{min}(E_R)}^\infty \frac{d^3\mathbf{v}_{rel}}{v_{rel}}f(\mathbf{v}_{rel}+\mathbf{v}_E),\notag\ \ \
    \xi(E_R)=\int_{v_{min}(E_R)}^\infty d^3\mathbf{v}_{rel}v_{rel}f(\mathbf{v}_{rel}+\mathbf{v}_E),
\end{align}
must be calculated when determining $\frac{dR}{dE_R}$. Analytical expressions for these integrals can be found in Appendix C of Ref.~\cite{Hisano:2018bpz}.}  
\begin{align}
    f_{\oplus}(\mathbf{v}_{rel})=f(\mathbf{v}_{rel}+\mathbf{v}_E).
\end{align}
For the numerical analysis we took the values used by the XENON1T collaboration \cite{Aprile:2018dbl}, namely, $v_0= 220$ km/s, $v_{\mathrm{esc}}=544$ km/s and $v_E=232$ km/s.

In the case of the direct detection experiment XENON1T, the number of expected events, $\mathcal{N}_{events}$, can be determined as \cite{Aprile:2011hx}
\begin{align}
    \mathcal{N}_{events}=\omega_{exp} \int_{S_{min}}^{S_{max}}dS\sum_{n=1}^\infty \text{Gauss}(S|n,\sqrt{n}\sigma_{PMT})\int_0^\infty dE_R\epsilon(E_R)\text{Poiss}(n|\nu(E_R))\frac{dR}{dE_R}.
\end{align}
Here $\omega_{exp}$ $=278.8$ days$\times$1.30 ton is the exposure, $S\in [3,70]$ is the number of photo-electrons (PE) resulting from the collision between the WIMP DM candidate and a Xe nucleus $\left(A=131\right.$, $Z=54$, $m_T=122.0 $ GeV$\left. \right)$; $\sigma_{PMT}$ is the average single-PE resolution of the photo-multipliers, $\epsilon(E_R)$ is the detection efficiency and $\nu(E_R)$ is the expected number of PEs for a given recoil energy $E_R$. For the numerical estimate of $\mathcal{N}_{events}$ we took $\sigma_{PMT}=0.4$ \cite{Aprile:2015lha, Barrow:2016doe}, whereas $\epsilon(E_R)$ was extracted from the black solid line in figure 1 of Ref.~\cite{Aprile:2018dbl}.  $\nu(E_R)$, for its part, was calculated as 
\begin{align}
    \nu(E_R)=E_R\ \mathcal{L}_{eff}\ \mathcal{L}_y\ S_{NR},
\end{align}
where the average light yield $\mathcal{L}y$ was fixed in 7.7 PE/keV \cite{Aprile:2015uzo} and a value of 0.95 was assigned to the light yield suppression factor for nuclear recoils $S_{NR}$. The relative scintillation efficiency $\mathcal{L}_{eff}$ was extracted from figure 1 in Ref.~\cite{Aprile:2011hi}.

From the most recent data reported by XENON1T, and with the aid of a Test Statistic (TS), we can obtain an upper bound for $\mathcal{N}_{events}$. Closely following Ref.~\cite{Cirelli:2013ufw}, we take
\begin{align}
    \text{TS}(m_\psi)=-2\ln \left[\frac{\mathcal{L}(\mathcal{N}_{events})}{\mathcal{L}_{BG}}\right],
\end{align}
with
\begin{align}
    \mathcal{L}(\mathcal{N}_{events}) = \frac{(\mathcal{N}_{events}+\mathcal{N}_{BG})^{\mathcal{N}_{obs}}}{\mathcal{N}_{obs}!}e^{-(\mathcal{N}_{events}+\mathcal{N}_{BG})},
\end{align}
and $\mathcal{L}_{BG}\equiv \mathcal{L}(0)$. It follows that by demanding $\text{TS}(m_\psi)>2.71$, limits for $\mathcal{N}_{events}$ are obtained at 90\% CL. For $\mathcal{N}_{obs}=14$ (number of observed events) and $\mathcal{N}_{BG}=7.36(61)$ (number of background events) \cite{Aprile:2018dbl}, the expected number of events must fulfill $\mathcal{N}_{events} \lesssim 19.5$~\cite{Hisano:2018bpz}.   

\section{Numerical results}
\label{sec:phenomenology}
\begin{table}[t]
\begin{center}
 \begin{tabular}{||c||}
  \hline \hline \rule[0cm]{0cm}{.25cm}
$ 1\,{\rm GeV}\leq m_{\psi_1}\leq 500\,{\rm GeV} $ \\ \hline \rule[0cm]{0cm}{.25cm}
$ 2\, m_{\psi_1}\leq m_{\psi_2}\leq 1\, {\rm TeV} $  \\ \hline \rule[0cm]{0cm}{.25cm}
$ {\rm max}\left(1.5\,m_{\psi_1},70 \,{\rm GeV}\right)\leq m_{S_1}\leq 1\,{\rm TeV}$  \\ \hline \rule[0cm]{0cm}{.25cm}
$ m_{S_1} \leq m_{S_2}\leq 2\,{\rm TeV} $  \\ \hline \rule[0cm]{0cm}{.25cm}
$ -\pi/2 \leq \varphi \leq \pi/2 $  \\ \hline \rule[0cm]{0cm}{.35cm}
$10^{-5}\leq |h_{\beta j}|\leq \sqrt{4\pi}$ \\ \hline \rule[0cm]{0cm}{.25cm}
$10^9\,{\rm GeV}\leq v_S \leq 10^{13}\,{\rm GeV}$ \\ \hline \rule[0cm]{0cm}{.25cm}
$0< \theta_a \leq 2\pi$ \\
\hline
 \end{tabular}
 \end{center}
 \caption{Random sampling for the relevant free parameters used in the numerical analysis.}
 \label{tab:parameterscan}
 \end{table}

In order to study the fermion DM phenomenology and take into account the constraints associated with charged LFV processes, we have implemented the model in {\tt SARAH} \cite{Staub:2013tta,Staub:2015kfa} to calculate, via {\tt SPheno} \cite{Porod:2003um,Porod:2011nf} and  {\tt FlavorKit} \cite{Porod:2014xia},  the flavor observables. 
In addition, we have used {\tt micrOMEGAs} \cite{Belanger:2013oya} to calculate the $\psi$ relic abundance. 
We have performed a random scan over the relevant free parameters of the model as shown in Table \ref{tab:parameterscan} and assumed $\lambda_2 = \lambda_{\Phi} = \lambda_{h\Phi} = \lambda_{S} = 0.01$. Moreover,  the mass of the exotic quark, $M_D$, has been set to $\sim 10$ TeV along with $f_{\beta} = 0.1$ in order to avoid the LHC constraints (see below). Let us recall that the $y_{i\beta}$ Yukawa couplings are linked to neutrino masses and the PMNS mixing matrix elements through the $h_{\beta j}$ Yukawa couplings as shown in section \ref{sec:model}. The LEP II constraints \cite{Achard:2001qw,Lundstrom:2008ai} on the $H^{\pm}$ charged scalar are automatically satisfied by the scan conditions defined in Table \ref{tab:parameterscan} and we also ensure that the oblique parameters $S$, $T$ and $U$ remain at $3\sigma$ level \cite{Baak:2014ora}\footnote{For simplicity purposes, we are taking $\lambda_3 = 0$ in such a way the charged and neutral components of the scalar doublet are degenerate.}. Concerning to the neutrino parameters, we consider both hierarchies for neutrino masses and use the best fit points values reported in Ref.~\cite{deSalas:2017kay} for the $\mathcal{CP}$ conserving case. Finally, regarding the charged LFV processes we consider the current experimental bounds and their future expectations as shown in table~\ref{tab:LFVprocessesexperimetal}.

\begin{table}[t!]
\begin{center}
\begin{tabular}{|cc|cc|c|}
\hline 
Observable & & Present limit & & Future sensitivity \\
\hline\hline \rule[0cm]{0cm}{.35cm}         
$\mathcal{B}(\mu \to e\gamma)$ & & $5.3 \times 10^{-13}$ \cite{Adam:2013mnn} & & $ 6.3 \times 10^{-14}$ \cite{Baldini:2013ke, MEGII:2018kmf} \\
$\mathcal{B}(\tau \to e\gamma)$&&$3.3 \times 10^{-8}$ \cite{Aubert:2009ag,Bona:2007qt,Miyazaki:2012mx}&& $3\times 10^{-9}$ \cite{Aushev:2010bq}\\
$\mathcal{B}(\tau\to\mu\gamma)$&& $4.4 \times 10^{-8}$ \cite{Aubert:2009ag,Bona:2007qt,Miyazaki:2012mx}&& $3\times 10^{-9}$ \cite{Aushev:2010bq}\\
$\mathcal{B}(\mu \to eee)$ & & $ 1.0\times 10^{-12}$ \cite{Bellgardt:1987du} & &$   10^{-16}$  \cite{Blondel:2013ia}   \\  
$\mathcal{B}(\tau \to eee)$ & & $ 4.4\times 10^{-8}$ \cite{Hayasaka:2010np} & &$  3\times 10^{-9}$  \cite{Aushev:2010bq}  \\  
$\mathcal{B}(\tau \to \mu\mu\mu)$ & & $ 2.1\times 10^{-8}$ \cite{Hayasaka:2010np} & &$   10^{-9}$  \cite{Aushev:2010bq}  \\  
${\rm R_{\mu e}({\rm Ti})}$ & & $4.3 \times 10^{-12}$ \cite{Dohmen:1993mp} & & $  10^{-18}$ \cite{Abrams:2012er} \\
${\rm R_{\mu e}({\rm Au})}$ & & $7.3 \times 10^{-13}$ \cite{Bertl:2006up} & &  $-$ \\
\hline
\end{tabular}
\end{center}
\caption{Current bounds and projected sensitivities for charged LFV observables.}
\label{tab:LFVprocessesexperimetal}
\end{table}

\begin{figure}[t]
  \includegraphics[scale=0.5]{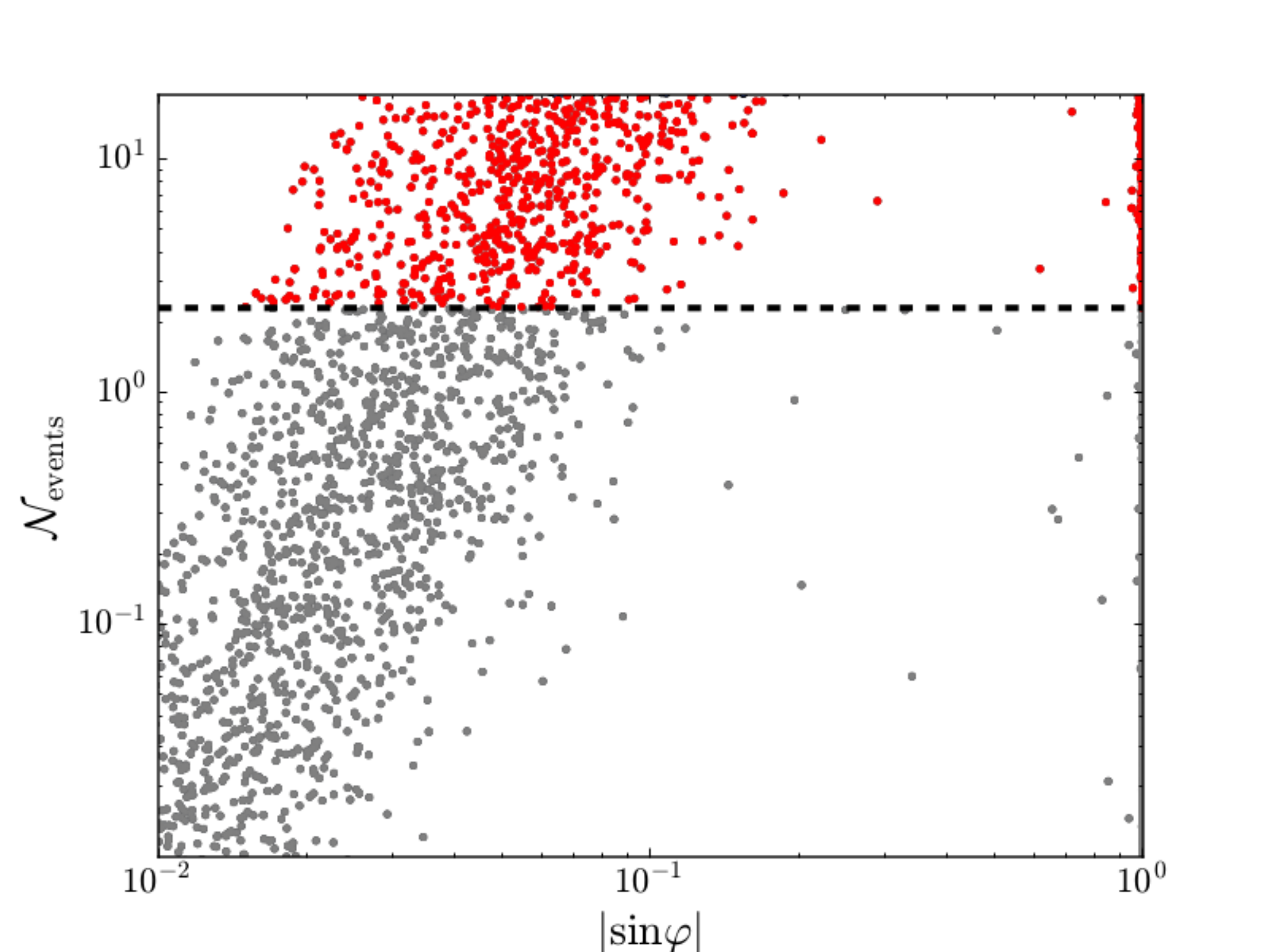}
  \caption{ Expected number of events as a function of the scalar mixing angle. All the points satisfy the current limit imposed XENON1T and the dashed line denotes the projected sensitivity of XENONnT experiment \cite{Aprile:2018dbl}.}
  \label{fig:Neventos}
\end{figure}

\begin{figure}[t!]
  \includegraphics[scale=0.5]{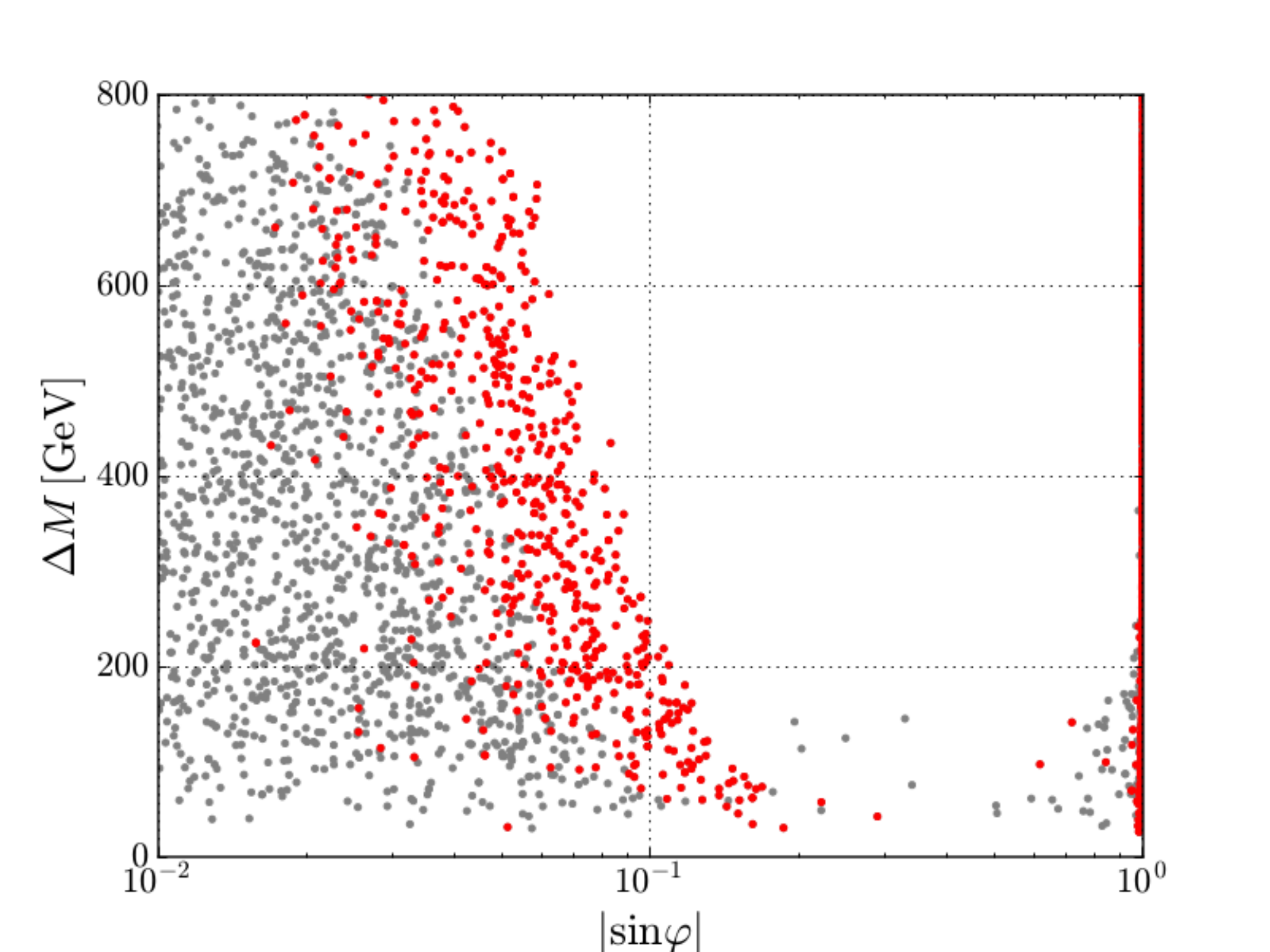}
  \caption{ The mass splitting $\Delta M\equiv m_{H^{\pm}} - m_{\psi}$ as a function of the scalar mixing angle. All the points satisfy the current limit imposed XENON1T and the red ones are those that will be explored by XENONnT. }
  \label{fig:Neventos-masas}
\end{figure}

We have calculated the fermion DM one-loop scattering cross section and 
estimated the expected number of events in direct detection experiments, $\mathcal{N}_{\text{events}}$, following the procedure described in Sec.~\ref{DirectDetection}. The results are shown in Figs. \ref{fig:Neventos} and \ref{fig:Neventos-masas}.
In Fig. \ref{fig:Neventos}, we present $\mathcal{N}_{\text{events}}$ as a function of the scalar mixing angle ($|\sin \varphi|$). All the points satisfy the current LFV constraints and the current limit imposed by the XENON1T collaboration. The prospect limits expected by XENONnT are indicated by the horizontal dashed line \cite{Aprile:2018dbl}. Notice that a large fraction of the parameter space (red points) will be explored in the next years (those featuring a small mixing angle are beyond the projected sensitivity, although such a small mixing angle is favoured by the tiny neutrino masses). 
From Fig. \ref{fig:Neventos} we also notice that for either $\varphi \rightarrow 0$ or $\varphi \rightarrow \pm\, \pi/2$, the number of events decreases rapidly, being steeper for $|\varphi| \to \pi/2$. This happens because when $\varphi \rightarrow 0$, the effective coupling between $\psi$ and the Higgs $\left(C_{\psi h}\right)$ becomes more suppressed than the case $|\varphi| \rightarrow \pi/2$ (see Eq. \eqref{Cpsih}). However, the coupling between $\psi$ and the Z boson ($C_{\psi Z}$) remains unaltered for both cases. 
Consequently, the expected number of events falls down more slowly for low small mixing angles.  
On the other hand, when the mixing is appreciable the contributions to $C_{\psi h}$ and $C_{\psi Z}$ coming from the neutral component of the scalar doublet $H_2$ 
become relevant, thus increasing the number of events in such a way it becomes maximum for $\varphi\sim \pi/4$. %

In Fig. \ref{fig:Neventos-masas} we display the mass splitting $\Delta M\equiv m_{H^{\pm}} - m_{\psi}$ as a function of the scalar mixing angle. As in Fig. \ref{fig:Neventos}, the gray points are allowed by the current experimental searches, whereas the red ones represent the region that will be explored in the next years. 
Notice that in the case of maximal mixing $\left(|\varphi| \sim \pi/4\right)$, where the number of events reaches its maximum value, there are some points localized in the allowed region. For these points the mass splitting between the charged scalar and $\psi$ is small, $\Delta M\lesssim 50$ GeV, which means that the co-annihilation processes are relevant. 
Conversely, for either $\varphi \rightarrow 0$ or $\varphi \rightarrow \pm\,\pi/2$ where the number of events drops sharply, the
mass splitting can take any value in the allowed range determined in the scan.
\begin{figure}[t!]
  \includegraphics[scale=0.5]{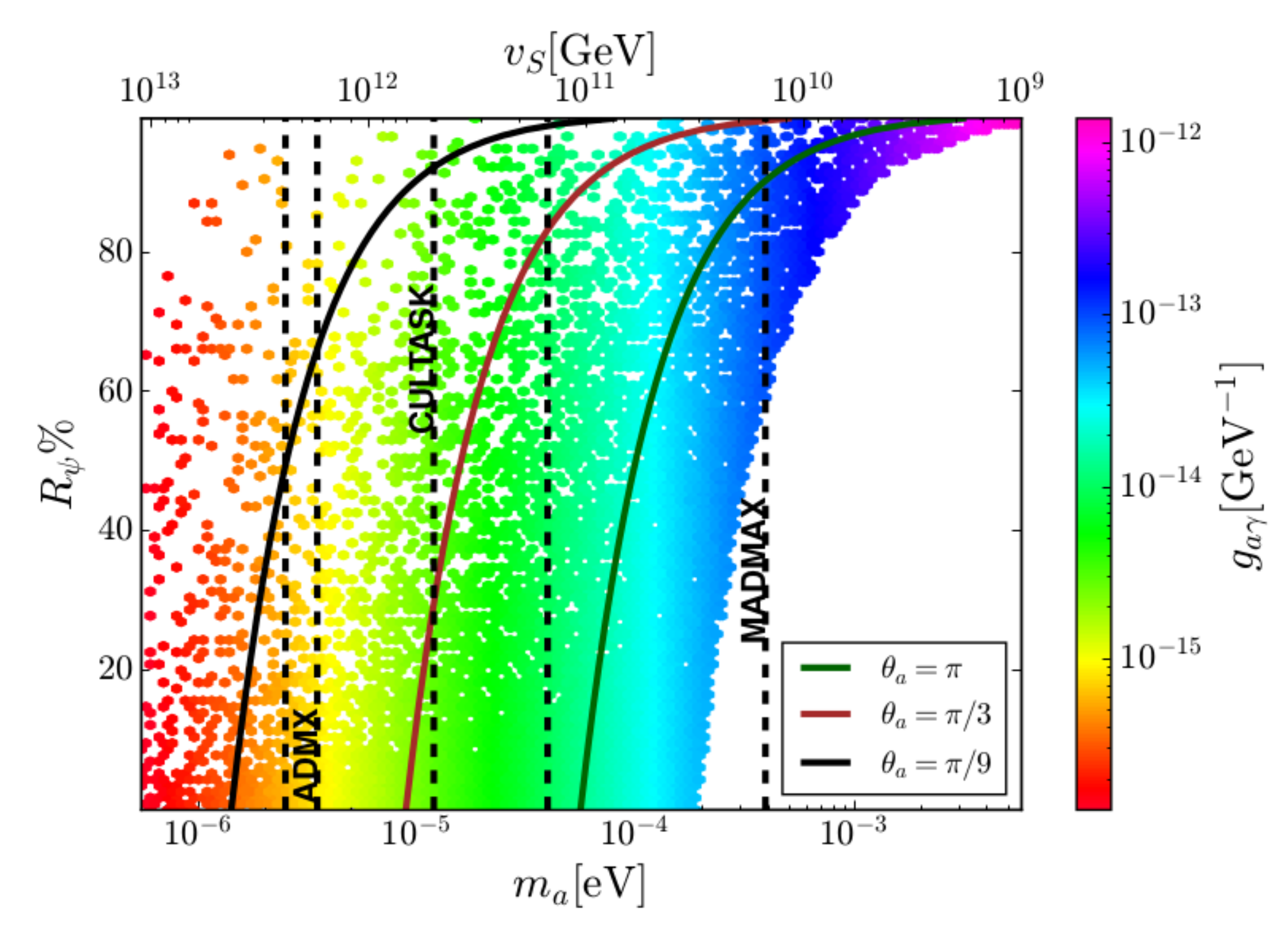}
  \caption{$\psi$ contribution to the total DM relic density as a function of the axion mass $m_a$ and the axion-photon coupling, $g_{a\gamma}\sim\frac{\alpha}{2\pi v_S}$, for different values of the initial misalignment angle $\theta_a$. 
  The vertical dashed lines represent the regions that will be explored by the cavity haloscope experiments ADMX \cite{ADMX:2018gho,Braine:2019fqb}, CULTASK \cite{Semertzidis:2019gkj, Chung:2018wms} and MADMAX \cite{Caldwell:2016dcw,MADMAX:2019pub,Millar:2016cjp,Horns:2012jf}.}
  \label{fig:axion_theta}
\end{figure}
We now turn to discuss the axion phenomenology. 
The contribution of $\psi$ to the total DM relic density $R_\psi=\Omega_\psi/(\Omega_\psi+\Omega_a)$ as a function of the axion mass is shown in Fig.~\ref{fig:axion_theta}.  
Each point reproduces the observed DM relic density $\Omega h^2  = 0.120 \pm 0.001 $ at $3\sigma$ \cite{Planck:2018vyg} and satisfies the direct detection constraints on $\psi$ as well as the charged LFV bounds. 
The color code represents the corresponding axion-photon coupling, $g_{a\gamma}\sim\frac{\alpha}{2\pi v_S}$, which plays a main role in axion searches through helioscope and haloscope experiments  \cite{Sikivie:2020zpn,Graham:2015ouw,Irastorza:2018dyq}. 
For $v_S\lesssim10^{10}$ GeV the main contribution to the DM relic abundance comes from the fermion DM candidate, with the corresponding axion mass window laying outside the experimental searches. However, for increasing values of $v_S$ the mixed fermion-axion DM scenario becomes more relevant and the axion can account for a fraction or the whole  of the DM relic abundance. 
In this case, a large fraction of the axion mass window (with the axion-photon coupling taking values in the range $g_{a\gamma}\sim(5\times10^{-16}-10^{-13})$ GeV$^{-1}$) can be explored by several haloscope experiments \cite{Graham:2015ouw,Irastorza:2018dyq}: ADMX for $m_a \sim (2.5-3.5)~ \mu$eV, CULTASK for $m_a \sim (3.5-12) ~\mu$eV and MADMAX for $m_a \sim(0.04-0.4)$ meV. 
Let us notice, however, that some regions are beyond the reach of the projected sensitivity of the experiments. 
Nonetheless, by enlarging the particle content or changing the PQ charge assignment on the current fields of the model, the chiral anomaly coefficient in the $g_{a\gamma}$ coupling can be modified in such a way that the entire region planned to search QCD axions in KSVZ and DFSZ models becomes experimentally accessible~\cite{Graham:2015ouw, Irastorza:2018dyq, DiLuzio:2017pfr}. 
\begin{figure}[t!]
  \includegraphics[scale=0.47]{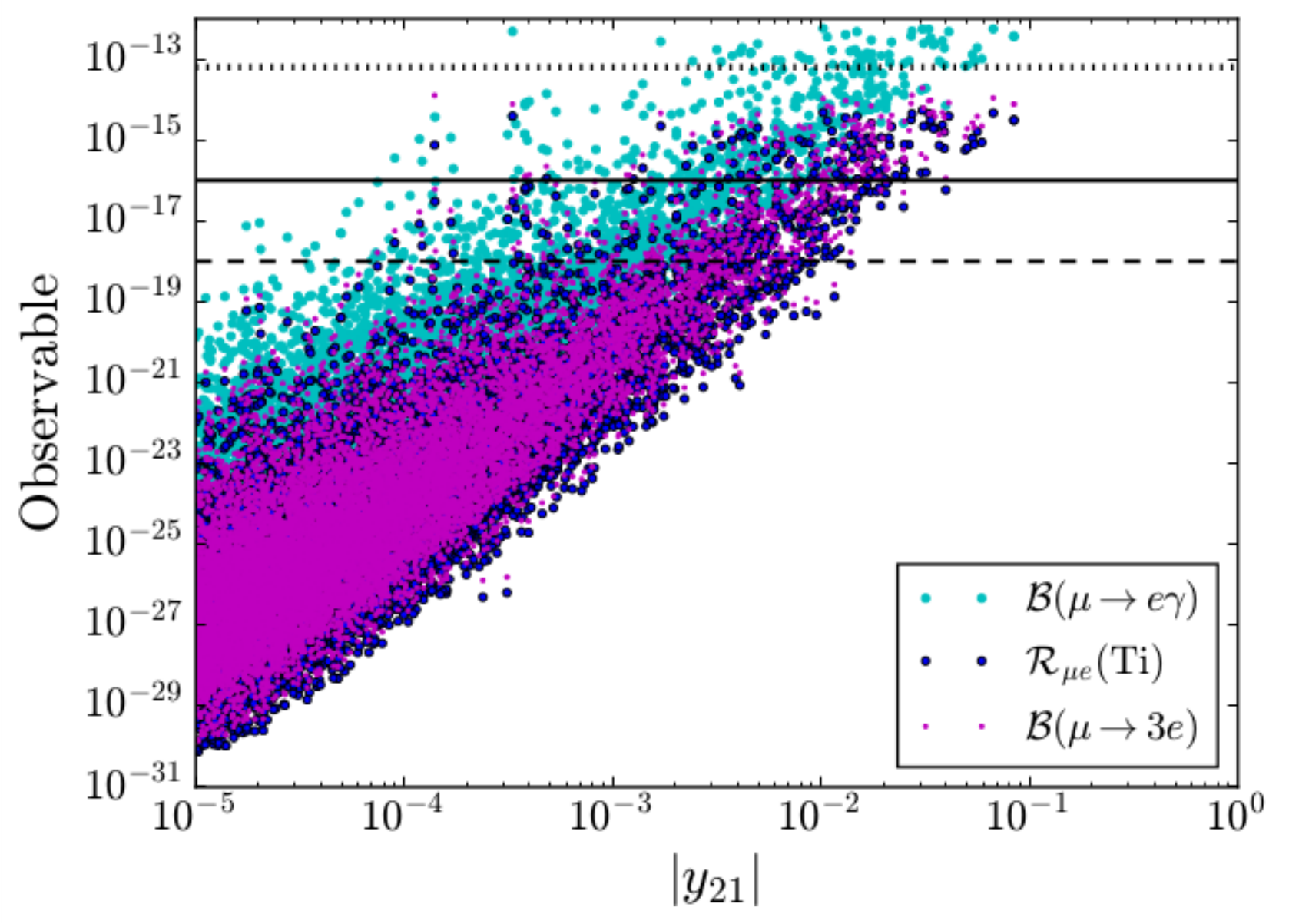}
  \includegraphics[scale=0.47]{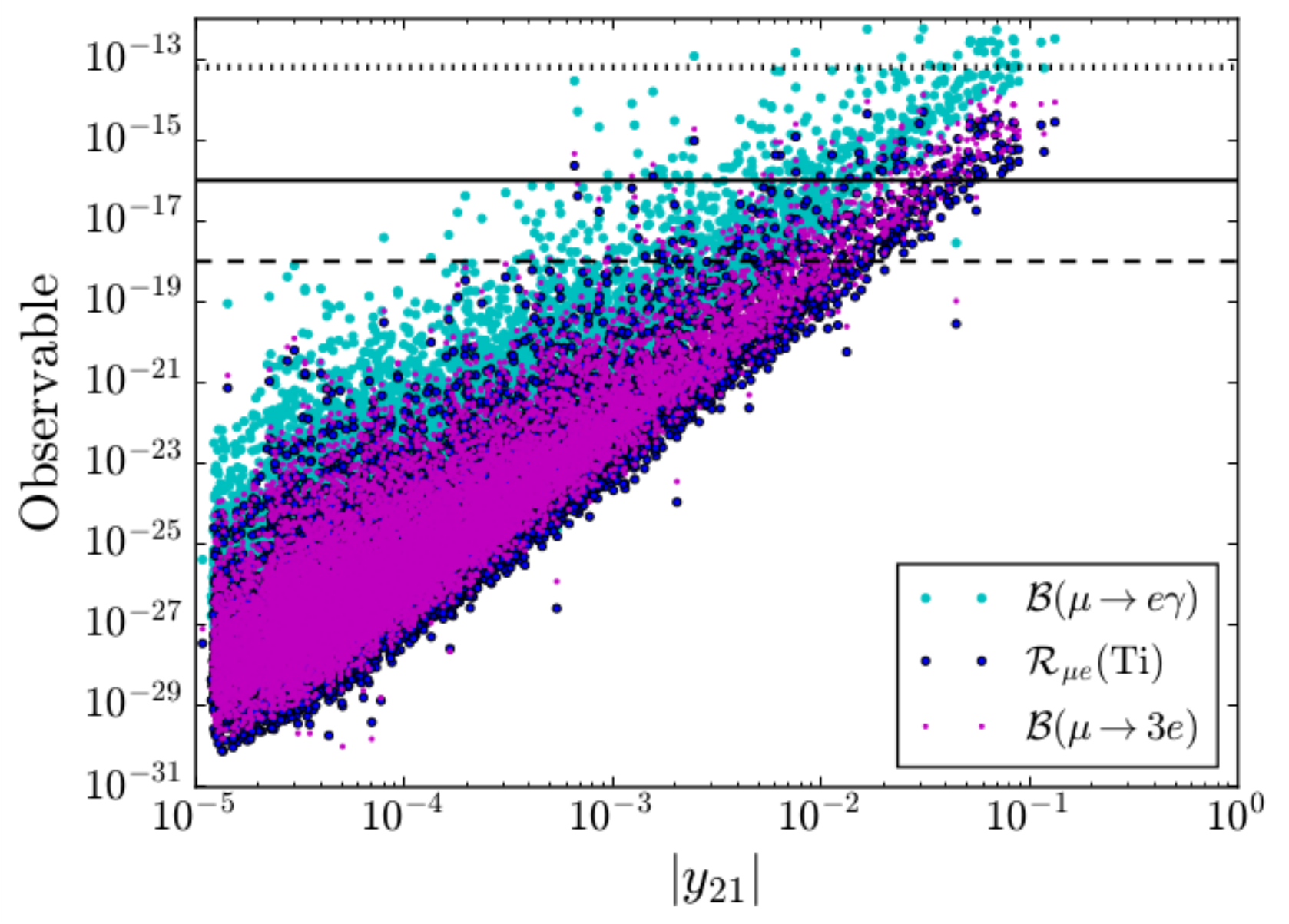}
  \caption{Branching ratios for charged LFV processes involving muons as a function of $|y_{21}|$ for the normal (left) and inverted (right) neutrino mass hierarchy. }
  \label{fig:LFVNHIH}
\end{figure}
\begin{figure}[t!]
  \includegraphics[scale=0.4]{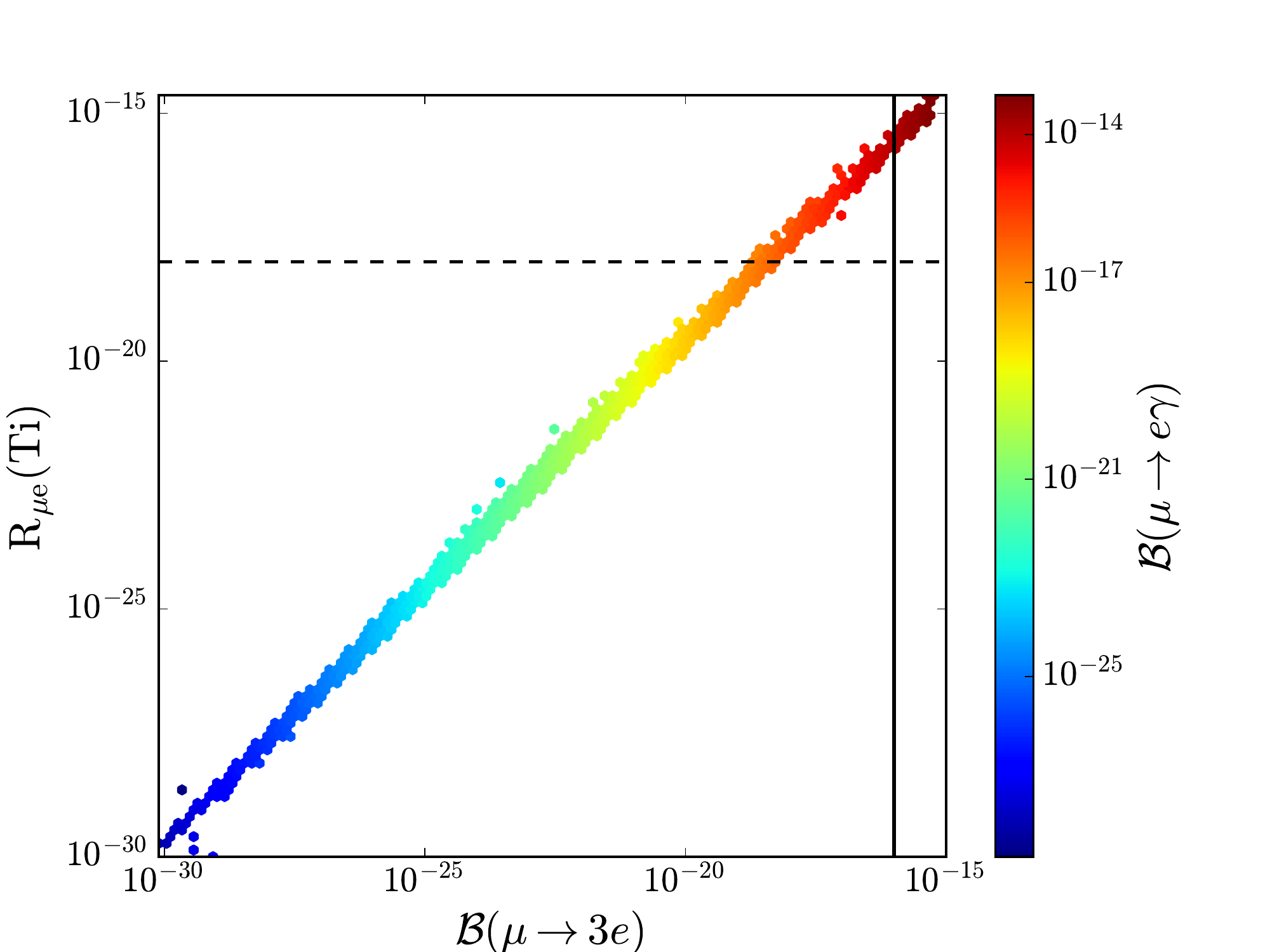}
   \includegraphics[scale=0.4]{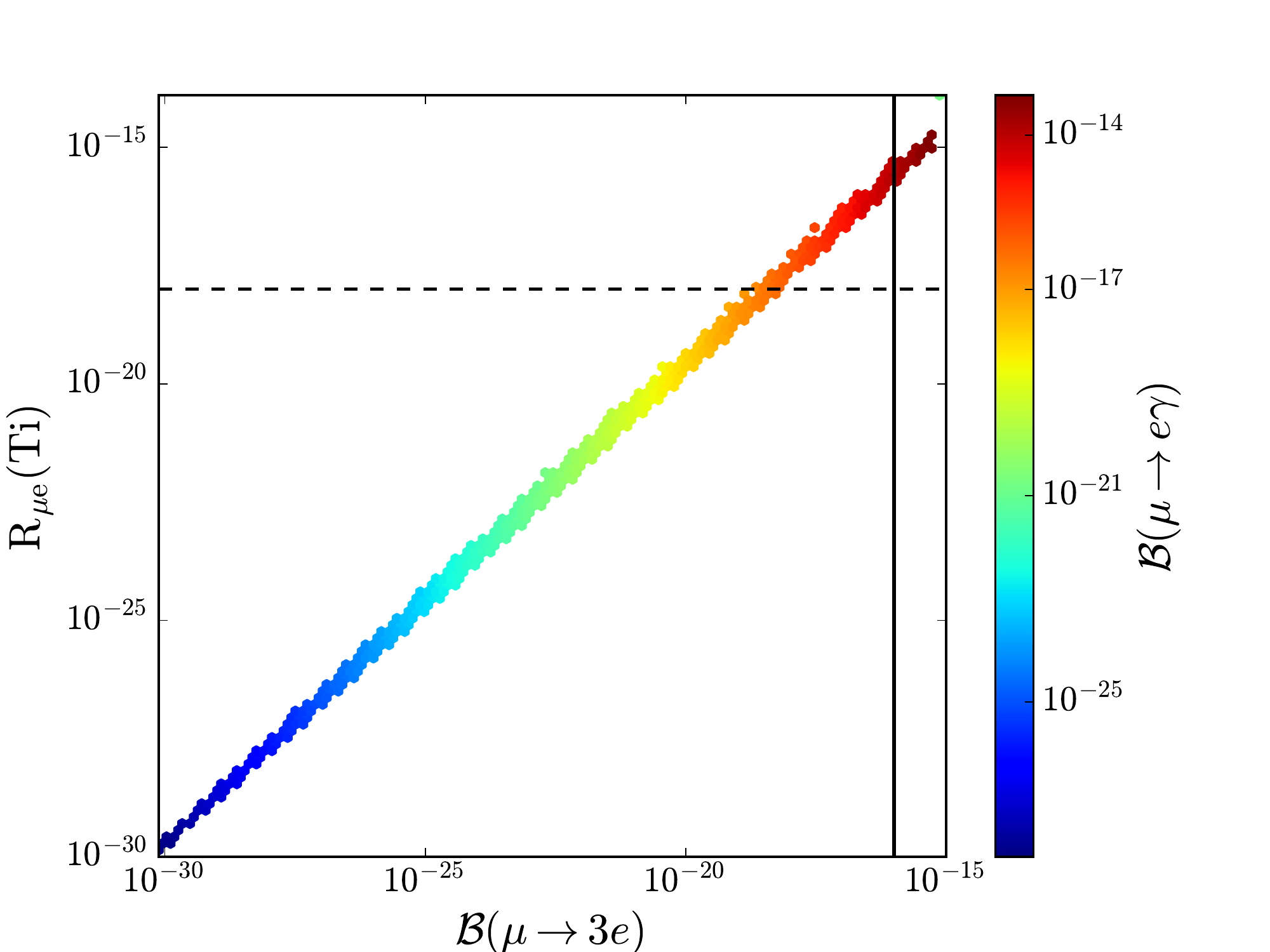}
  \caption{Correlation between the different  LFV observables involving muons. The left and right panels correspond to a IH and NH, respectively.  The solid and dashed lines represent the expected sensitivity for the future searches $\mathcal{B}(\mu\rightarrow 3e)$ and $\mathcal{R}_{\mu e}$, respectively.}
  \label{fig:futureLFV}
\end{figure}

Regarding the charged LFV processes, we focus in the observables involving muons in the initial state. In Fig.~\ref{fig:LFVNHIH} are displayed the branching ratios  $\mathcal{B}(\mu \rightarrow e\gamma)$ (cyan points), $\mathcal{B}(\mu \rightarrow 3e)$ (magenta points) and the rate for the $\mu \rightarrow e$ conversion in titanium $\mathcal{R}_{\mu e}$  (blue points) as a function of the Yukawa coupling $y_{21}$. The left (right) panel stands for the normal (inverted) neutrino mass hierarchy, with the dotted, solid and dashed horizontal lines representing the projected sensitivity of LFV experiments for $\mathcal{B}(\mu\rightarrow e \gamma)$, $\mathcal{B}(\mu\rightarrow 3e)$ and $\mathcal{R}_{\mu e}$, respectively. 
It follows that for both neutrino mass hierarchies the current bounds demand $|y_{21}|\lesssim 0.1$ whereas the future searches will explore values as low as 0.01, with the conversion in nuclei being the most relevant process (we have found similar results for the other Yukawa couplings $y_{i\beta}$).   
In addition to this, we notice that the observables associated with the lepton conversion in nuclei and the process with three electrons in the final state are strongly correlated. This is because the conversion process in nuclei does not involve box diagrams whereas the corresponding contribution in $\mu \rightarrow 3e$ becomes suppressed by a factor $|y_{i\beta}|^4$ and therefore the penguin diagrams give the dominant contribution for both charged LFV processes. This strong correlation, along with the correlation from $\mathcal{B}(\mu \rightarrow e\gamma)$, is shown in Fig. \ref{fig:futureLFV}. 
On the other hand, we observed that the Yukawa couplings $h_{\beta i}$ can take values along the whole range considered in the scan. This result along with the ones for $y_{i\beta}$ discussed above translate to that the scalar coupling $|\lambda'|$ lies below $10^{-7}$ in order to reproduce the observed neutrino mass scale (see Eq.~\eqref{neutrinomasses2}).

A final comment is in order concerning the exotic quark $D$. Since it couples to the SM sector through the Yukawa term, $\overline{q_L}H_2D_R$, it can decay into a $S_{1,2}$ scalar and a SM quark, thus avoiding potential issues that arise when an exotic quark is considered cosmologically stable \cite{Nardi:1990ku}. Furthermore, such an exotic  quark can be produced at colliders via quark and gluon fusion, leading to the specific signature of jets plus missing energy. From the analysis presented in Ref.~\cite{Alves:2016bib}, which deals with a scenario similar to the one studied here, the LHC searches for exotic quarks imply that $M_D \gtrsim $ 700 GeV, which is well below the value considered in our analysis. 

\section{Conclusions}
\label{sec:summary}
In the class of models known as scotogenic models the generation of radiative neutrino masses is associated with the existence of a discrete symmetry that forbids the tree-level contribution and stabilizes the DM particle.  On these lines,  the PQ symmetry can also be invoked to simultaneously provide a solution to the strong CP problem, radiatively induce neutrino masses and stabilize the particles lying in the dark sector. 
In this work we considered a multicomponent scotogenic model with Dirac neutrinos where the dark sector is composed by the QCD axion and a SM singlet fermion, the latter stabilized by the PQ symmetry.     
We computed the expected number of fermion-DM nuclei scatterings in XENON1T and identified regions of the parameter space compatible with observed DM abundance, direct searches of singlet fermion and axion DM, neutrino oscillation data and the upper bounds on charged LFV processes. 
Furthermore,  we find that for some choices of parameters both the singlet fermion and the axion present detection rates are within the expected sensitivity of XENONnT and haloscope experiments such as ADMX, CULTASK, and MADMAX, respectively, and that the LFV processes involving muons in the initial state may be probed in upcoming rare leptonic decay experiments. 

\section{Acknowledgments}
We are thankful to Diego Restrepo and Nicolás Bernal for enlightening discussions.  
This work has been supported by Sostenibilidad-UdeA and the UdeA/CODI Grants 2017-16286 and 2020-33177. Cristian D. R. Carvajal and Robinson Longas  acknowledge the financial support given by COLCIENCIAS through the doctoral scholarship 727-2015 and 617-2013.

\appendix 

\section{One-loop functions for effective interactions of fermion DM} \label{sec:oneloopfunctions}
In this section we reproduce the expressions for the one-loop functions required to calculate the $C^\gamma_M$, $C^\gamma_R$, $C_{\psi h}$ and $C_{\psi Z}$ effective couplings \cite{Hisano:2018bpz}.  
The functions associated with $C_M^\gamma$, $C_R^\gamma$ and $C_{\psi h}$ and are given by 
\begin{align}
g_{M1}(m_\psi, M, m) &= 1-\frac{\M-\m}{2\mx}\mathrm{ln}\left(\frac{\M}{\m}\right)+\frac{\Delta+\mx\left(\M-\mx+\m\right)}{2\mx}L ,\label{gM1}\\
g_{R1}(m_\psi, M, m) &= \frac{1}{12}\left[\frac{8\left(\M-\m\right)+\mx}{\mx}\mathrm{ln}\left(\frac{\M}{\m}\right)-\frac{4}{\Delta}\left\lbrace 4\Delta +\mx\left(\M+3\m\right)-m^4_\psi\right\rbrace\right.\nonumber\\
&\left.-\frac{L}{\mx\Delta}\left\lbrace 8\Delta^2 + \left(9\M-5\mx+7\m\right)\mx \Delta - 4\m m^4_\psi \left(3\M-\mx+\m\right)\right\rbrace\right],\label{gR1}
\end{align}
whereas, the one associated with $C_{\psi h}$ reads
\begin{align}
\tilde{g}_{h1}\left(m_\psi,M_i,M_j,m\right) &= \frac{1}{2}+\frac{\Mi+\Mj-2\left(\m+\mx\right)}{4\mx}\mathrm{ln}\left(\frac{\m}{M_i M_j}\right)\notag\\
&-\frac{M^4_i+M^4_j-2\left(\Mi+\Mj\right)\left(\m+\mx\right)-2\left(m^4_{\psi}-m^4\right)}{4\mx\left(\Mi-\Mj\right)}\mathrm{ln}\left(\frac{M_i}{M_j}\right)\notag\\
&+\frac{(\Mi-\m-\mx)\Delta_i L_i-(\Mj-\m-\mx)\Delta_j L_j}{4\mx(\Mi-\Mj)}.
\end{align}
The computation of $C_{\psi Z}$ involves the functions
\begin{align}
 g_{Z1}(m_\psi,M,m_i,m_j)&=-\frac{1}{2}\Delta_{\epsilon,\mu}+\frac{1}{2}\mathrm{ln}\ M+\frac{\mi\ \mathrm{ln}\ m_i - \mj\ \mathrm{ln}\ m_j}{2\left(\mi-\mj\right)}+\frac{\mi+\mj-2\M-\mx}{2\mx}\notag\\
&+\dfrac{\mathrm{ln}\left(\frac{\M}{m_i m_j}\right)}{4m^4_\psi}\left[m^4_i+m^4_j-\mi\mj-3\M\left(\mi +\mj-\M\right)\right. \notag\\
&\left. -\mx\left(\mi+\mj+2\M\right)\right] -\frac{\mathrm{ln}\left(\frac{m_i}{m_j}\right)}{4m^4_\psi\left(\mi-\mj\right)}\left\lbrace 2m^6_\psi-6m^4_\psi\M+\mx \left[ 6M^4\right.\right.\notag\\
&\left.\left. -2M^2\left(\mi+\mj\right)-m^4_i-m^4_j\right] -2M^6+3M^4\left(\mi+\mj\right)-3\M\left(m^4_i+m^4_j\right)\right.\notag\\
&\left. +m^6_i+m^6_j\right\rbrace+\frac{1}{4m^4_\psi\left(\mi-\mj\right)}\left\lbrace\left[\left(M^2-\mi\right)^2+\left(M^2-\mx\right)^2-M^4\right]\Delta_i L_i\right. \notag\\ &\left.-\left[\left(M^2-\mj\right)^2+\left(M^2-\mx\right)^2-M^4\right]\Delta_j L_j\right\rbrace,
\end{align}
\begin{align}
g_{Z2}(m_\psi, M, m_i, m_j)& = \frac{1}{2}\mathrm{ln}\left(\frac{M^2}{m_i m_j}\right) +\frac{\mathrm{ln}\left(\frac{m_i}{m_j}\right)}{2\left(\mi-\mj\right)}\left(2M^2-2\mx-\mi-\mj\right)\nonumber\\
&+\frac{\Delta_iL_i-\Delta_jL_j}{2\left(\mi-\mj\right)},
\end{align}
\begin{align}
\tilde{g}_{Z1}(m_\psi,M_i,M_j,m)&=\frac{1}{2}\Delta_{\epsilon,\mu}-\frac{1}{2}\mathrm{ln}\ m -\frac{\Mi\ \mathrm{ln}\ M_i - \Mj\ \mathrm{ln}\ M_j}{2\left(\Mi-\Mj\right)}+\frac{\Mi+\Mj-2\m+\mx}{2\mx}\notag\\
&+\frac{3\m\left(\m-\Mi-\Mj\right)-\mx\left(\Mi+\Mj\right)+M^4_i+M^4_j+\Mi\Mj}{4m^4_\psi}\mathrm{ln}\left(\frac{\m}{M_iM_j}\right)\notag\\
&-\frac{\mathrm{ln}\left(\frac{M_i}{M_j}\right)}{4m^4_\psi\left(\Mi-\Mj\right)}\left[M^6_i+M^6_j-\left(3\m+\mx\right)\left(M^4_i+M^4_j\right)+3m^4\left(\Mi+\Mj\right)\right.\notag\\
&\left.-2\left(\m-\mx\right)^2\left(\m+\mx\right)\right]+\frac{1}{4m^4_\psi\left(\Mi-\Mj\right)}\left\lbrace\left[\left(\Mi - \m\right)^2-m^4_\psi\right]\Delta_i L_i\right.\notag\\ 
&\left.- \left[\left(\Mj - \m\right)^2-m^4_\psi\right]\Delta_j L_j\right\rbrace,
\end{align}
where
\begin{align}
\Delta\left(m_\psi, M, m\right)&= m^4_\psi - 2\mx\left(\M+\m\right)+\left(\M-\m\right)^2 \notag\\
&= m^4 - 2\m \left(\M + \mx\right) + \left(\M -\mx\right)^2 \notag\\
&= M^4 - 2\M \left(\mx +\m\right)+\left(\mx - \m\right)^2 ,\label{Delta}
\end{align}
and \footnote{This expression for $L$ is valid provided that $\Delta > 0$, which is always true in our model.}
\begin{align}
L\left(m_\psi, M, m\right) = \frac{1}{\sqrt{\Delta}}\mathrm{ln}\left(\frac{M^2+m^2-\mx+\sqrt{\Delta}}{M^2+m^2-\mx-\sqrt{\Delta}}\right) . \label{L}
\end{align}
Additionally, $\Delta_i$ and $L_i$ denote the quantities $\Delta(m_\psi, M_i, m)$ and $L(m_\psi, M_i, m)$, respectively, while $\Delta_{\epsilon,\mu}$ is a divergent constant term.

From these general expressions we deduce appropriated formulas for the limiting cases $m\to0$, $M_i = M_j$ and $m_i = m_j$. 
In what follows, and when possible, the resulting expressions will be written in terms of the parameters $x \equiv \dfrac{m}{M}$, $y \equiv \dfrac{m_\psi}{M}$ and $\delta \equiv x^4-2x^2(1+y^2)+(1-y^2)^2$. Thus, for example,
\begin{align}
g_{M1}(m_\psi,M,m) &= 1+\frac{1-x^2}{y^2}\mathrm{ln}\ x + \frac{(1-x^2)^2-(1+x^2)y^2}{2y^2\sqrt{\delta}}\mathrm{ln}\left(\frac{1+x^2-y^2+\sqrt{\delta}}{1+x^2-y^2-\sqrt{\delta}}\right) ,\\
g_{R1}(m_\psi,M,m) &= -\frac{1}{6}\left[ 8\frac{1-x^2}{y^2}+1\right]\mathrm{ln}\ x -\frac{1}{3\delta}\left[4(1-x^2)^2-(7+5x^2)y^2+3y^4\right]\notag\\
&-\frac{1}{12y^2\delta^{3/2}}\mathrm{ln}\left(\frac{1+x^2-y^2+\sqrt{\delta}}{1+x^2-y^2-\sqrt{\delta}}\right)\left[8(1-x^2)^4-(1-x^2)^2(23+25x^2)y^2\right.\notag\\
&\left.+(25-2x^2+25x^4)y^4-(13+11x^2)y^6+3y^8\right] .
\end{align}
For the case $M_i \neq M_j$ and $m = 0$ we obtain
\begin{align}
\tilde{g}_{h1}(m_\psi,M_i,M_j,0)&=\frac{1}{2} + \frac{m^2_\psi}{M^2_i-M^2_j}\mathrm{ln}\left(\frac{M_i}{M_j}\right)\notag\\
&+\frac{(M^2_i-m^2_\psi)^2\ \mathrm{ln}\left(\dfrac{M_i^2-m^2_\psi}{M^2_i}\right)-(M^2_j-m^2_\psi)^2\ \mathrm{ln}\left(\dfrac{M_j^2-m^2_\psi}{M^2_j}\right)}{2m^2_\psi(M^2_i-M^2_j)},\\
\tilde{g}_{Z1}(m_\psi,M_i,M_j,0)&= \frac{1}{2}\Delta_{\epsilon,\mu}+\frac{1}{2m^4_\psi(M^2_i-M^2_j)}\left[(M^4_i-m^4_\psi)(M^2_i-m^2_\psi)\ \mathrm{ln}\left(\dfrac{M_i^2-m^2_\psi}{M^2_i}\right)\right.\notag\\
&\left.-(M^4_j-m^4_\psi)(M^2_j-m^2_\psi)\ \mathrm{ln}\left(\dfrac{M_j^2-m^2_\psi}{M^2_j}\right)\right]+\frac{M^2_i+M^2_j+m^2_\psi}{2m^2_\psi},
\end{align}
whereas for $M_i = M_j\equiv M$ and $m \neq 0$ 
\begin{align}
\tilde{g}_{h1}(m_\psi,M,M,m)&=1+\frac{1-x^2-y^2}{y^2}\mathrm{ln}\ x+\frac{x^4-2x^2+(1-y^2)^2}{2y^2\sqrt{\delta}}\mathrm{ln}\left(\frac{1+x^2-y^2+\sqrt{\delta}}{1+x^2-y^2-\sqrt{\delta}}\right),\\
\tilde{g}_{Z1}(m_\psi,M,M,m)&= \frac{1}{2}\Delta_{\epsilon,\mu}-\frac{1}{2}\mathrm{ln}(mM)+\frac{1}{4}+\frac{6-6x^2-y^2}{4y^2}+\frac{3-2y^2-3x^2(2-x^2)}{2y^4}\mathrm{ln}\ x\notag\\
&+\frac{1}{4y^4\sqrt{\delta}}\mathrm{ln}\left(\frac{1+x^2-y^2+\sqrt{\delta}}{1+x^2-y^2-\sqrt{\delta}}\right)\left[3(1-x^2)^3+(3x^4+2x^2-5)y^2\right.\notag\\
&\left.+(1-x^2)y^4+y^6\right],
\label{gZ1t}
\end{align}
The case $M_i = M_j\equiv M$ and $m = 0$ leads to  
\begin{align}
\tilde{g}_{h1}(m_\psi,M,M,0)&=1+\frac{1-y^2}{y^2}\mathrm{ln}\left(1-y^2\right)\ ,\label{gh1mMM0}\\
\tilde{g}_{Z1}(m_\psi,M,M,0) &= \frac{1}{2}\Delta_{\epsilon,\mu} -\mathrm{ln}M + \frac{1}{4}-\frac{y^2-6}{4y^2} + \frac{3-5y^2+y^4+y^6}{2y^4(1-y^2)}\mathrm{ln}(1-y^2)\ ,
\end{align}
and the case $m_i = m_j\equiv m \neq 0$   involves the limiting functions
\begin{align}
g_{Z1}(m_\psi,M,m,m)&= -\frac{1}{2}\Delta_{\epsilon,\mu}+\frac{1}{2}\mathrm{ln}(mM)-\frac{1}{4}-\frac{x^4+3(1-2x^2)-2y^2(1+x^2)}{2y^4}\mathrm{ln}\ x\notag\\ 
&-\frac{3(1-x^2)^3-5(1-x^4)y^2+(1-3x^2)y^4+y^6}{4y^4\sqrt{\delta}}\mathrm{ln}\left(\frac{1+x^2-y^2+\sqrt{\delta}}{1+x^2-y^2-\sqrt{\delta}}\right)\notag\\
&+\frac{6x^2+y^2-6}{4y^2},
\label{gZ1} \\
g_{Z2}(m_\psi,M,m,m)&= -\mathrm{ln}\ x-\frac{1-x^2+y^2}{2\sqrt{\delta}}\mathrm{ln}\left(\frac{1+x^2-y^2+\sqrt{\delta}}{1+x^2-y^2-\sqrt{\delta}}\right).
\end{align}
Finally, the limiting functions appearing in the case $m_i = m_j\equiv m = 0$ fulfill
\begin{align}
g_{Z1}(m_\psi,M,0,0) &= -\tilde{g}_{Z1}(m_\psi,M,M,0).\label{gZ10}
\end{align}
Notice that, from Eqs. \eqref{gZ1t} and \eqref{gZ1}
\begin{align}
g_{Z1}(m_\psi,M,m,m) + \tilde{g}_{Z1}(m_\psi,M,M,m) &= \frac{(x^2+y^2)}{x^{-2}y^4}\mathrm{ln}\ x \nonumber\\ &+\frac{(1-x^2+y^2)}{2x^{-2}y^2\sqrt{\delta}}\mathrm{ln}\left(\frac{1+x^2-y^2+\sqrt{\delta}}{1+x^2-y^2-\sqrt{\delta}}\right).
\end{align}

\bibliographystyle{apsrev4-1long}
\bibliography{references}

\end{document}